\documentclass[preprint,superscriptaddress,showpacs,aps,pre,floatfix,longbibliography]
{revtex4-1}
\usepackage[utf8]{inputenc} 
\usepackage[T1]{fontenc}    
\usepackage{booktabs}       
\usepackage{amsfonts}       
\usepackage{nicefrac}       
\usepackage{microtype}      
\usepackage{amssymb,amsthm}
\usepackage{amsmath,bm}
\usepackage{graphicx,epstopdf}
\usepackage{multirow}
\usepackage{subfigure}
\usepackage{setspace}
\usepackage{wrapfig}
\usepackage{comment}
\usepackage{listings}
\usepackage{amssymb,dsfont}
\usepackage{leftidx}
\usepackage{bigdelim}
\usepackage{mathrsfs}
\usepackage{blkarray,adjustbox}
\usepackage{lineno}
\usepackage{enumitem}
\usepackage{mathpazo}
\usepackage{rotating}
\usepackage{soul}
\usepackage[dvipsnames]{xcolor}
\usepackage{xr-hyper}
\usepackage{xspace}
\usepackage{float}
\usepackage[pdftex, pdftitle={Article},colorlinks, pdfauthor={Author}]{hyperref} 
\usepackage[nameinlink,capitalise]{cleveref}
\hypersetup{colorlinks={true},linkcolor={darkgray},citecolor=ForestGreen}

\definecolor{reddish}{HTML}{FBB4AE}
\definecolor{blueish}{HTML}{B3CDE3}
\definecolor{magentish}{HTML}{FF00AA}
\definecolor{greenish}{HTML}{a1d99b}

\begin{document}
\title{Contrasting and comparing the efficacy of non-pharmaceutical interventions on air-borne and vector-borne diseases}

 \author{Bibandhan Poudyal}
      \affiliation{Department of Physics \& Astronomy, University of Rochester, Rochester, NY, 14627, USA}
\author{David Soriano Pan\~os}
 \email[Correspondence email address: ]{sorianopanos@gmail.com}
\affiliation{Departament d'Enginyeria Inform\`atica i Matem\`atiques, Universitat Rovira i Virgili, 43007 Tarragona (Spain)}
\affiliation{GOTHAM lab, Institute for Biocomputation and Physics of Complex Systems (BIFI), University of Zaragoza, 50018 Zaragoza (Spain)}
 \author{Gourab Ghoshal}
     \email[Correspondence email address: ]{gghoshal@pas.rochester.edu}
 \affiliation{Department of Physics \& Astronomy, University of Rochester, Rochester, NY, 14627, USA}
 \affiliation{Department of Computer Science, University of Rochester, Rochester, NY, 14627, USA}

\begin{abstract}
Non-pharmaceutical interventions (NPIs) aimed at limiting human mobility have demonstrated success in curbing the transmission of airborne diseases. However, their effectiveness in managing vector-borne diseases remains less clear. In this study, we introduce a framework that integrates mobility data with vulnerability matrices to evaluate the differential impacts of mobility-based NPIs on both airborne and vector-borne pathogens. Focusing on the city of Santiago de Cali in Colombia, our analysis illustrates how mobility-based policies previously proposed to contain airborne disease can make cities more prone to the spread of vector-borne diseases. By proposing a simplified synthetic model, we explain the limitations of the latter policies and exploit the synergies between both types of diseases to find new interventions reshaping the mobility network for their simultaneous control. Our results thus offer valuable insights into the epidemiological trade-offs of concurrent disease management, providing a foundation for the design and assessment of targeted interventions that reshape human mobility.
\end{abstract}

\maketitle

\section{Introduction}
The role of mobility in the propagation of epidemics is well-established~\cite{barbosa2018human}. On the one hand, the rapid expansion of global mobility networks has dramatically increased the level of inter-connectivity between regions, accelerating the transmission of infectious diseases~\cite{baker2022infectious,nowzari2016analysis,wang2013effect}. On the other hand, the ongoing trends in urbanization and the transition toward more densely populated urban environments further exacerbate the potential for localized disease outbreaks to escalate into epidemics~\cite{connolly2021extended,sharifi2020covid,wu2017economic}.  Accurately predicting and mitigating these outbreaks requires a comprehensive understanding of epidemic vulnerability, encompassing factors such as mobility patterns, population susceptibility, healthcare infrastructure, and environmental conditions~\cite{adger2009nested,brockmann2009human,world2014vector,suk2014interconnected,massaro2019assessing,barbosa2021uncovering, lee2017morphology, Mimar_2022, Poudayal_2023}.

In the absence of universally effective vaccines and therapeutics, non-pharmaceutical interventions (NPIs) have become a critical tool in mitigating the spread of epidemics~\cite{lison2023effectiveness,perra2021non,markel2007nonpharmaceutical}. Nonetheless, the suitability of NPIs as optimal responses to epidemic threats hinges on the balance between their expected health benefits and the collateral effects derived from their implementation~\cite{angulo2021simple,correia2022pandemics,pangallo2024unequal}. For instance, policies shaping human mobility during the COVID-19 pandemic were efficient in controlling the exponential growth of cases yet introduced significant socioeconomic challenges~\cite{GDPandem24:online, muller2015annual, GrossDom47:online,Valganon:2023rw}.

Beyond their associated socioeconomic cost, the implementation of NPIs tailored to mitigate a specific epidemic scenario affects the epidemic trajectories of other diseases circulating in the same population. Such effects are always synergistic when both diseases share the same transmission mechanisms, as proven by the reduced prevalence of airborne diseases (ABDs) in the population resulting from the implementation of NPIs designed to contain COVID-19 cases~\cite{feng2021impact,huang2021impact,fricke2021impact}. Conversely, quantifying the impact of such policies on other circulating pathogens with different transmission mechanisms, such as vector-borne diseases (VBDs), represents a more intricate problem. Indeed, several empirical studies have reported disparate results of these interventions ranging from potential benefits to negligible effects or even unintended drawbacks \cite{brady2021impact, ong2021implications, cavany2021pandemic, chen2022measuring, sharma2022does, lu2021dengue, surendran2022reduced, widyantoro2021implication, Zimbabwe1:online, rodriguez2023vector,khan2021prioritizing}. 

Numerous theoretical studies have investigated the impact of local mobility on the global transmission of both ABDs and VBDs \cite{baker2022infectious,pastor2015epidemic,soriano2022modeling,gomez2018critical,castillo2016perspectives,soriano2020vector,massaro2019assessing,Aguilar:2022kl}. However, a holistic understanding of the combined spatial and temporal dynamics of mobility on disease spread remains incomplete \cite{balcan2009multiscale,belik2011natural,balcan2010modeling,hancean2020impact,soriano2022modeling,castillo2016perspectives,poletto2013human,blonder2012temporal,holme2016temporal,ruan2007spatial}. In this study, we present a novel and unified framework for analyzing both vector-borne diseases (VBDs) and airborne diseases (ABDs) at a large scale. Our approach is grounded in a metapopulation model \cite{soriano2020vector,soriano2022modeling,hazarie2021interplay,lloyd2004spatiotemporal} designed to capture the complexity of multi-patch systems. This model incorporates a matrix-based representation of epidemic vulnerability, with each row representing the overall vulnerability of a particular patch.  Utilizing the LouBar method \cite{li2018effect,bassolas2019hierarchical}, we classify regions into high-vulnerability zones ("hotspots") and low-vulnerability zones ("suburbs") based on their population densities. This allows us to identify key areas and relevant mobility flows that shape cities' vulnerabilities to both types of diseases.

Using this framework and real data from the city of Cali in Colombia, we first demonstrate that interventions proven successful in reducing ABD transmission do not yield comparable outcomes for VBDs. We explain this disparity, by exploring the contagion dynamics of both disease types on a simplifed one-hub-one-leaf mobility network, disentangling the temporal and spatiotemporal components. This approach enables the identification of optimal conditions for the simultaneous control of both ABDs and VBDs in a region. Leveraging this information, we redesign the interventions on the mobility network of the city of Cali, showing that model-informed policies are efficient in containing ABDs while not hampering the control of VBDs. The successful implementation of these strategies using real-world metapopulation data underscores the potential of the model to inform public health interventions for both airborne and vector-borne diseases.

\section{Methods}
\subsection{Data}

As a primary illustration, we examine the spread of vector-borne diseases (VBDs) and airborne diseases (ABDs) within Cali, Colombia. This metropolitan area, with a population exceeding 2 million, provides an ideal test case due to its documented history of severe VBD outbreaks. To assess the impact of mobility patterns on disease transmission within Cali, a detailed reconstruction of the city's resident mobility network was conducted. This reconstruction involved dividing the city into 22 designated administrative districts, or "comunas." Demographic information on population distribution across these comunas was sourced from the local municipality's census records.
Subsequently, mobility flows between these districts were derived from established urban commuting surveys \cite{escobar2013cali}, resulting in the collection of over $10^5$ individual travel trajectories, which offer a substantial representation of commuting patterns within Cali. This dataset was then used to construct an origin-destination matrix, $\textbf{R}$ whose elements correspond to $R_{ij} = T_{ij}/ \sum_{j} T_{ij}$, where $T_{ij}$ encodes the population outflows from comuna $i$ to $j$. The elements of $\textbf{R}$ can thus be interpreted as the probability of moving from $i$ to $j$. In Fig. S1{\bf a} we show the resulting mobility matrix ${\bf R}$ and in Fig. S1{\bf b} we show the mobility flows on the map of Cali. 

To model the vector distribution (mosquito populations) across comunas, we consider the {\em recipient index}, encoding the probability of finding a mosquito in different recipients distributed across the city. We thus assume that a high recipient index corresponds with a large concentration of vectors in this area. Following this rationale, we assume that the ratio between the number of vectors and humans inside each patch in our model is directly proportional to its recipient index, which is extracted from entomological data of the year 2015~\cite{secretaria2016cali}. Table~\ref{tab:summary_entomology} presents summary statistics for entomological indices in the 22 communas of Cali, Colombia.
\begin{table*}[t!]
\renewcommand{\arraystretch}{1.7} 
\centering 
\begin{tabular}{|c|c|c|c|c|}
\hline
\textbf{Count} & \textbf{Mean} & \textbf{SD} & \textbf{Min} & \textbf{Max}\\
\hline
22 & 1.38 & 0.52 & 0.6 & 2.7\\
\hline
\end{tabular}
\caption{Summary statistic for the recipient index distribution in Cali. This quantifies the percentage of recipients found with larvae and pupae across the city of Cali. The number of mosquitoes in each area is computed assuming that the ratio of vectors to humans within each patch is proportional to this index.}
\label{tab:summary_entomology}
\end{table*}

\subsection{Hotspot classification}

Hotspots are identified using the LouBar method, where hotspots are determined by setting a threshold based on population densities. First, patches are arranged in ascending order of density. A Lorenz curve is then plotted, with the cumulative density of patches on the Y-axis and corresponding patch numbers on the X-axis. The derivative of the curve at its peak point (patch with maximum cumulative density) is extrapolated to intersect the X-axis, determining the threshold patch beyond which all patches are considered hotspots. A patch $i$ qualifies as a hotspot if $\rho_i > \rho^{LouBar}$, where $\rho_i$ is the density of patch $i$ and $\rho^{LouBar}$ is the threshold density obtained via the LouBar method. All remaining patches are referred to as \emph{suburbs}. For further details on this method, see Fig. S2 where we plot the hotspots in terms of the population density (upper-panel) as well as the \emph{effective density} (ratio of mosquitoes to human population in a given comune). 

\subsection{Epidemic vulnerability}

The concept of an epidemic threshold $\tau_c$ is frequently used to determine the minimum level of infectiousness necessary for a pathogen to establish within a population. A higher epidemic threshold signifies increased resistance to disease transmission, whereas a lower threshold suggests heightened vulnerability. In this study, we define epidemic vulnerability as the inverse of the epidemic threshold. Thus, a lower epidemic threshold, indicative of greater susceptibility to disease transmission, equates to a higher level of epidemic vulnerability. Consequently, the epidemic vulnerability, $\nu = \tau_c ^{-1}$, for a given geographic area (referred to as a "patch") is defined as the expected number of contagions the population within that area encounters over time, primarily influenced by the population's recurrent mobility patterns.

To assess the city's susceptibility to airborne diseases (ABDs) and vector-borne diseases (VBDs) independently, we employ a metapopulation framework. Building on established models for ABDs \cite{hazarie2021interplay} and VBDs \cite{soriano2020vector}, we adapt these frameworks to suit the specific context of this study. The dynamics of ABDs are represented using a generalized Susceptible-Infected-Recovered (SIR) model, while the Ross-Macdonald (RM) model is applied to describe VBD transmission~\cite{RMD_2012}. Both models incorporate spatial distributions and mobility patterns, fundamental components of metapopulation theory. In this framework, a metapopulation is conceptualized as a network where nodes represent geographic locations (patches) and edges denote population movement between these patches.

Each patch $i$ is characterized by its population size $n_i$, area $a_i$, and vector population $m_i$. These attributes vary across patches, reflecting differences in demographic distribution and vector prevalence. Human and vector populations are assigned to specific patches based on their residence locations, with $n_i$ and $m_i$ representing the number of humans and vectors residing in patch $i$. The patches are interconnected, forming a directed, and weighted network encoded in the mobility matrix $\textbf{R}$. This matrix governs the movement of the human population across the system, while mosquitoes are assumed to remain confined to their respective patches due to their limited dispersal range~\cite{muir1998aedes}. For further details on the model, refer to Supplementary Section S3, and see Table~\ref{tab:definitions} for a list of key parameters.
\begin{table*}[t!]
\renewcommand{\arraystretch}{1.5}
\begin{tabular}{ | p{1cm}| p{14cm}|}
 \hline
 \multicolumn{2}{|c|}{\textbf{Parameter table}} \\
 \hline
\textbf{} & \textbf{Definition}\\
 \hline
 $\alpha$ &  Ratio of human population in the leaf to human population in the hub. 
\\
\hline
 $\beta$ &  Ratio of mosquito population in the leaf to mosquito population in the hub. 
\\
\hline
 $\gamma$ &  Ratio of area of the leaf to area of the hub .
\\
\hline
 $\kappa$ &  Fraction of population moving within hub. 
\\
\hline
 $1-\kappa$ &  Fraction of population moving from hub to leaf. 
\\
\hline
 $\delta$ & Fraction of population moving within leaf. 
\\
\hline
 $1-\delta$ &  Fraction of population moving from leaf to hub. 
\\
\hline
\end{tabular}
\caption{Definitions of the parameters used in this paper.} 
\label{tab:definitions}
\end{table*}


\section{Results}
\subsection{Effect of NPIs in Cali, Colombia}
 \begin{figure}[t]
    \centering
    \includegraphics[width=0.8\columnwidth]{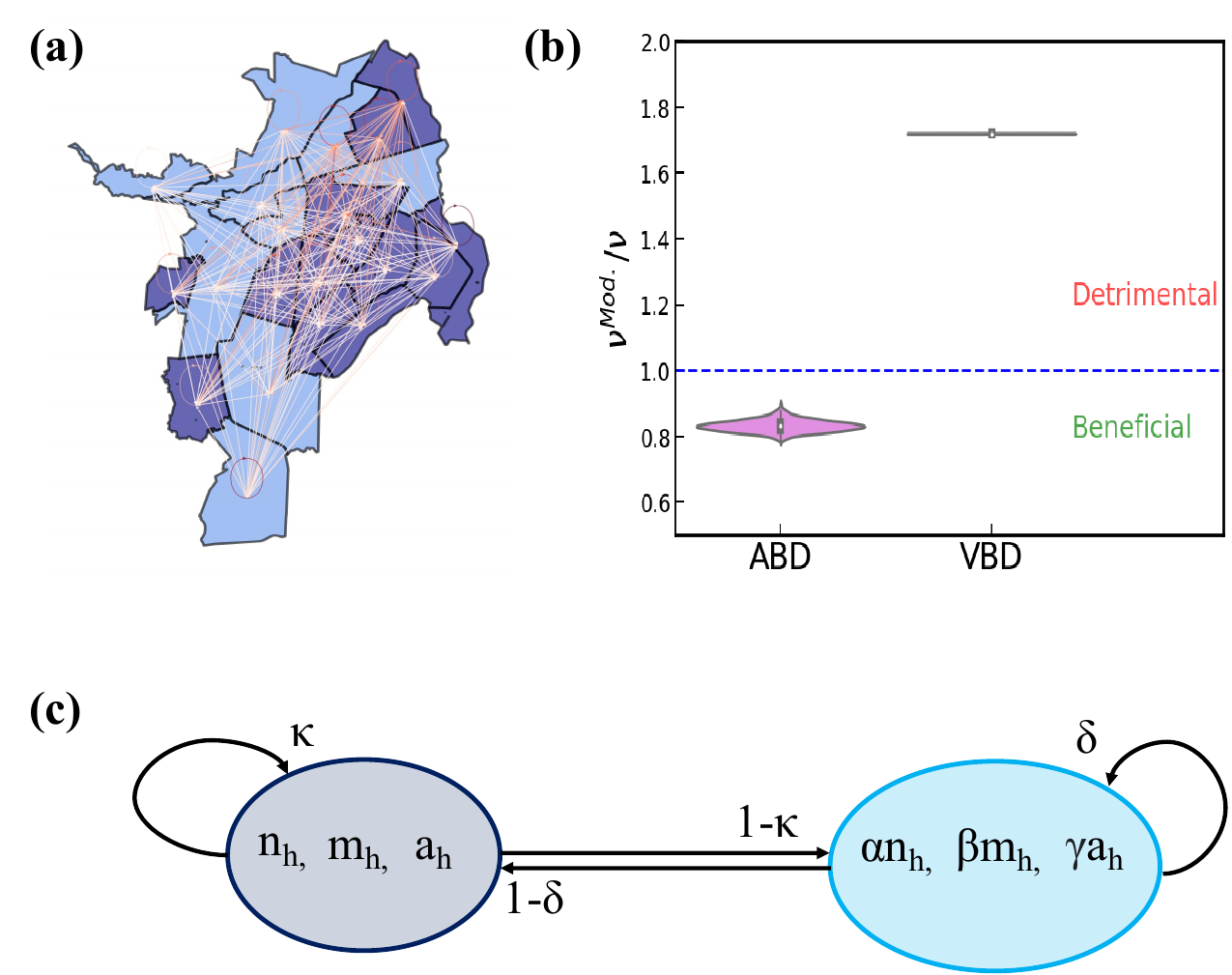}
    \caption{\textbf{Effect of NPIs in Cali and a Simplified Metapopulation Model.} {\bf (a)} Map of Cali showing the spatial distribution of identified population density hotspots (navy) and suburbs (light blue), along with the mobility network, where edge weights are represented using a gradient red color scale.
{\bf (b)} Illustration of a Non-Pharmaceutical Intervention (NPI) involving the redistribution of mobility flows from hotspots to neighboring suburbs. The effect of the intervention on epidemic vulnerability is quantified by the ratio $\frac{\nu^{Mod}}{\nu}$, where $\nu^{Mod}$ represents the vulnerability recalculated post-intervention. Values below 1 indicate a beneficial intervention, reducing vulnerability, while values above 1 signify a detrimental effect, increasing vulnerability.
{\bf (c)} Schematic of the simplified one hub-leaf metapopulation model. The hub aggregates all hotspots and is defined by its human population ($n_h$), mosquito population ($m_h$), and area ($a_h$). The leaf, encompassing all non-hotspot regions (suburbs), is characterized by its human population ($n_l = \alpha n_h$), mosquito population ($m_l = \beta m_h$), and area ($a_l = \gamma a_h$), with $\alpha$, $\beta$, and $\gamma$ acting as scaling parameters that adjust the relative sizes of the hub and leaf.}
    \label{fig:figure1}
\end{figure}

In Fig.\ref{fig:figure1}\textbf{a}, we present a map of Cali, highlighting hotspot comunas in red. For reference, the population density distribution across comunas is provided in Fig. S2. One effective (albeit somewhat idealized) NPI for reducing regional vulnerability to ABDs, as proposed in\cite{hazarie2021interplay}, involves rerouting mobility flows from hotspots to neighboring suburban areas while maintaining the total outflow from each hotspot. For each realization of this modified flow structure, we compute the resulting vulnerability, $v^{mod}$. In Fig.~\ref{fig:figure1}\textbf{b}, we show the distribution of the ratio of modified vulnerability (across 1,000 iterations) to the baseline vulnerability calculated from Cali's empirical mobility network. As anticipated, this intervention achieves an average reduction in regional vulnerability of approximately 20\%. This positive effect is attributed to the population density asymmetry within cities; by diverting mobility from dense hotspots to lower-density suburbs, the intervention raises the epidemic threshold through a dilution effect, reducing potential contacts within hotspots and distributing potentially infectious individuals to areas where their impact is minimized.

Conversely, this approach has the unintended consequence of worsening conditions for VBDs. The hotspots defined by vector densities (e.g., mosquito populations) generally differ from those based on human population densities, although a positive correlation exists between the two (Spearman correlation coefficient of 0.74; see Fig. S2). When computing modified vulnerabilities for VBDs under the same intervention, we observe a near doubling of vulnerability. The fact that an NPI strategy effective for ABDs may not yield similar results for VBDs emphasizes the necessity of tailoring epidemic preparedness strategies to the specific characteristics of each disease. 

\subsection{Synthetic hub-leaf metapopulation}

Our finding---that an intervention effective for one disease type may exacerbate the vulnerability of another within the same region---emphasizes the necessity of thoroughly examining the contagion dynamics of ABDs and VBDs and their interactions with population density and mobility flows. In the case of Santiago de Cali, this interplay could be further investigated by analyzing the relationship between epidemic vulnerability and the structural architecture of its mobility network. However, the inherent complexity of human mobility patterns poses challenges, specifically: (i) extracting actionable insights directly from mobility network analyses, and (ii) generalizing these findings to inform disease control strategies in other urban settings.

To address these challenges, we adopt an alternative approach by focusing on universal factors that constrain the effectiveness of control policies, using a simplified synthetic model. This model reduces the complexity of mobility networks by coarse-graining them into a two-patch metapopulation, consisting of a single hub and a single leaf. The simplification assumes that all nodes classified within the same category---either hotspots or suburbs, as identified by the LouBar method---are equivalent in terms of their underlying contagion dynamics (See Figure S3 for a schematic of the formalism).

The primary features of the one hub-leaf metapopulation are illustrated in Fig.~\ref{fig:figure1}{\bf c}. The hub is characterized by its resident population $n_h$, area $a_h$, and vector population $m_h$. Similarly, the leaf is defined by three corresponding parameters: $n_l$, $a_l$, and $m_l$. To facilitate the analysis, we relate the parameters of the leaf to those of the hub using the relationships $n_l = \alpha n_h$, $a_l = \gamma a_h$, and $m_l = \beta m_h$, where $\alpha$, $\gamma$, and $\beta$ are scaling factors that adjust the ratios of human population, area, and mosquito population, respectively, between the two patches.

Mobility dynamics in this model are governed by probabilistic movement: individuals from the hub move to the leaf with probability $\kappa$ or remain in the hub with probability $1-\kappa$, while individuals from the leaf move to the hub with probability $\delta$ or remain in the leaf with probability $1-\delta$. This framework allows us to systematically explore the dependence of urban vulnerability on the parameter space defined by $(\alpha, \gamma, \beta, \kappa, \delta)$, facilitating the identification of universal policies that minimize vulnerability across diverse urban environments.

\subsection{Contagion dynamics in ABDs}

We begin by analyzing the well-established mechanisms governing airborne disease transmission assuming recurrent mobility in our synthetic model. Building on this framework (see Supplementary Sections S3-S4 for further details), the vulnerability $\nu$ of a simplified one-hub-leaf model for ABDs can be expressed as 
\begin{align}\label{eq:equation1}
    \nu^{ABD} = \frac{\Theta_h(\kappa,\delta) + \frac{\Theta_l(\kappa,\delta)}{\gamma} + \sqrt{\left(\Theta_h(\kappa,\delta) + \frac{\Theta_l(\kappa,\delta)}{\gamma} \right)^2 - \frac{\alpha}{\gamma}\left(\Theta_{hl}(\kappa,\delta)\right)^2}}{2},
\end{align}
where $\Theta_h(\kappa,\delta)$, $\Theta_l(\kappa,\delta)$ and $\Theta_{hl}(\kappa,\delta)$ can be interpreted as hub-, leaf- and cross-vulnerabilities respectively. Their precise mathematical forms are provided in Supplementary Materials Eq. S26.
It is straightfoward to show that $\nu^{ABD}$ is minimized when $\kappa \approx \frac{1}{\gamma + 1}$ and $\delta \approx \frac{\gamma}{\gamma + 1}$.

For our analysis, we simplify the model by fixing $\gamma = 1$, ensuring that the hub and leaf regions have equal areas. This approach eliminates the confounding effects of spatial heterogeneity, allowing us to focus on the contagion dynamics. Under this configuration, a homogeneous distribution of the population across the hub and leaf is optimal. Achieving such a distribution requires selecting mobility parameters that promote an even population spread between the regions. Crucially, the initial population ratio between the leaf and hub, represented by $\alpha$, emerges as a key determinant in this process. (For a detailed analysis of this limiting case and a breakdown of each term in Eq.~\eqref{eq:equation1}, cf. Suplementary material Secs.~S4, S5.)

\begin{figure}[t]
    \centering
    \includegraphics[width=16cm]{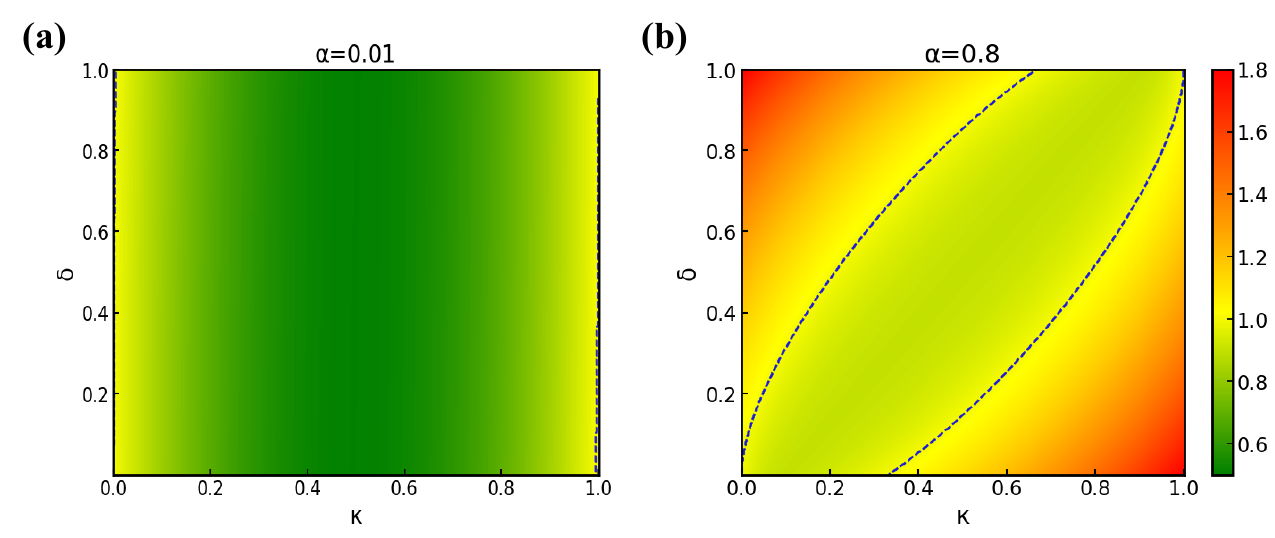}
    \caption{\textbf{Epidemic vulnerability in ABDs.} \textbf{(a)} and \textbf{(b)}: Contour plots illustrating the dynamics of vulnerability in airborne disease (ABD) for different values of the parameter $\alpha$. The blue dotted contour line indicates the threshold where vulnerability equals 1. In both plots, $\gamma=1$.}
    \label{fig:figure2}
\end{figure}


In Fig.~\ref{fig:figure2}\textbf{a}, we illustrate the vulnerability dynamics for a scenario where $\alpha$ is set to a small value, representing a relatively low population in the leaf compared to the hub. In this regime, the hub exhibits the highest vulnerability, and the system's behavior is predominantly governed by the mobility parameter $\kappa$. For low leaf populations, migration from the hub to the leaf is critical to achieving a more balanced distribution. As $\kappa$ increases, a greater proportion of the hub's population migrates to the leaf, reducing vulnerability within the hub. When $\kappa \approx 0.5$, roughly half of the hub population has moved to the leaf, resulting in a near-homogeneous distribution across the two regions. However, as $\kappa \to 1$, nearly the entire hub population migrates to the leaf, significantly increasing the leaf's vulnerability and establishing it as the new hotspot.


In epidemiological models, dilution refers to the process of homogenizing population distribution by redistributing individuals across regions. In the context of our ABD model, larger values of $\alpha$, which correspond to larger leaf populations, reduces the extent of dilution. This reduced dilution elevates overall vulnerability, as reflected by the more intense red hue in the color scale of Fig.~\ref{fig:figure2}\textbf{b}. When the leaf-to-hub population ratio approaches 1, both the leaf and hub exhibit equal susceptibility to vulnerability. In this case, any migration from the hub to the leaf requires a proportional counterflow from the leaf to the hub to achieve optimal population homogenization across the two regions.

In Fig.~\ref{fig:figure2}\textbf{b}, vulnerability is minimized along the upward diagonal, where $\kappa \approx \delta$. This configuration represents a balanced migration regime, where the movement of individuals from the hub to the leaf is counterbalanced by an equivalent flow from the leaf to the hub. Deviations from this equilibrium lead to imbalanced population distributions and increased vulnerability. Specifically, lower $\kappa$ and higher $\delta$ result in an accumulation of individuals in the leaf, causing a sharp increase in its vulnerability. Conversely, higher $\kappa$ and lower $\delta$ lead to a concentration of individuals in the hub, similarly raising its vulnerability. Thus, extreme values of mobility in either direction result in higher overall system vulnerability. In Supplementary Materials, Fig. S4, we plot the analog of Fig.\ref{fig:figure2} for other intermediate values in the range $0 \leq \alpha \leq 1$ mirroring the trends described here. 

\subsection{Contagion dynamics in VBDs}

\begin{figure}[t!]
    \centering
    \includegraphics[width=16cm]{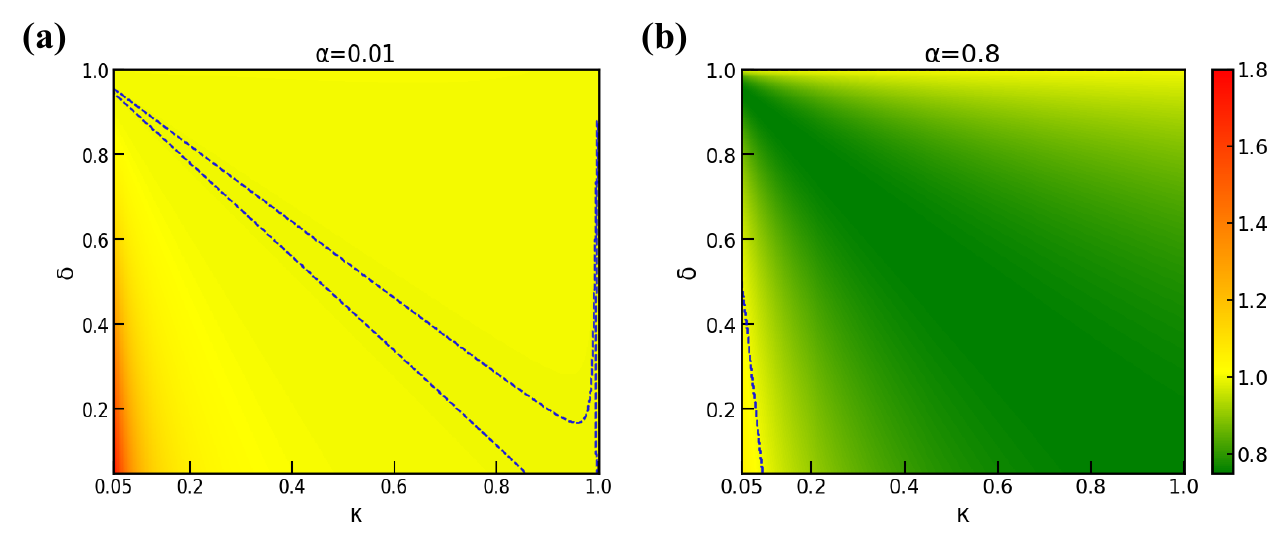}
    \caption{\textbf{Epidemic vulnerability in VBDs} \textbf{(a)} and \textbf{(b)}: Contour plots illustrating the dynamics of vulnerability in vector-borne disease (VBD) for different values of the parameter $\alpha$. The blue dotted contour line indicates the threshold where vulnerability equals 1. In both plots, $\beta=0.01$.}
    \label{fig:3}\
\end{figure}

Next, we analyze the dynamics specific to VBDs. In this context, the vulnerability $\nu^{VBD}$ represents the average number of indirect contacts an individual has with the rest of the population, mediated by vectors. The corresponding equation for the vulnerability (Details in Supplementary Sections S1 and S2), is:
\begin{align}\label{eq:equation2}
    \nu^{VBD} = \sqrt{\frac{\Tilde{\Theta}_h(\kappa,\delta) + \beta\Tilde{\Theta}_l(\kappa,\delta) + \sqrt{\left(\Tilde{\Theta}_h(\kappa,\delta) + \beta\Tilde{\Theta}_l(\kappa,\delta) \right)^2 - \alpha\beta\left(\Tilde{\Theta}_{hl}(\kappa,\delta)\right)^2}}{2},}
\end{align}
where once again the terms retain their interpretation as in ABDs, with their precise mathematical forms shown in Supplementary Materials Eq. S30.
We note now the appearance of the term $\beta$ which quantifies the relative population of mosquitoes in the leaf as compared to the hub $m_l/m_h$.
The term $\beta m_i$ represents the total number of interactions occurring within patch $i$, independent of the number of individuals occupying the patch. However, larger effective populations reduce the likelihood of any specific individual being bitten during these interactions, which explains the presence of effective populations in the denominator of the associated expressions. For our subsequent analysis, motivated by the observed correspondence of mosquito- and population density hubs in Fig. S2, we simplify the model by fixing $\beta = 0.01$, ensuring a higher mosquito population in the hub region, which designates it as a potential hotspot. This configuration allows for an apples-to-apples comparison with the ABD model, where hotspots are similarly characterized by higher densities in their respective critical factors.

In the VBD model, smaller values of $\alpha$, corresponding to lower leaf populations, reduce the scale of dilution, thereby increasing overall vulnerability. This is reflected by the more intense red hue in the color scale of Fig.~\ref{fig:3}\textbf{a}. In this regime, the hub and leaf exhibit similar vulnerabilities in the absence of mobility, i.e., when $(\delta, \kappa) = (1, 1)$. As with ABDs, starting from this scenario and reducing $\kappa$ increases the hub's vulnerability, as emptying the hub heightens its local exposure. However, mobility also allows hub residents to leave the area, which offsets the increase in local vulnerability, resulting in an approximately constant value of $\nu = 1$ along this axis.
Similarly, altering $\delta$ to increase the movement of individuals from the leaf reduces the time residents spend in the leaf while increasing its role as a center of transmission. Conversely, when $\kappa$ approaches 0 and $\delta$ is low, the absence of hub residents in the hub amplifies the risk of exposure for the smaller incoming leaf population, leading to heightened overall vulnerability.

For higher values of $\alpha$, the vulnerabilities of the hub and the leaf in the absence of mobility differ significantly, as $\beta / \alpha \ll 1$. This discrepancy makes the beneficial effects of mobility, primarily through the dilution of hub vulnerability, more pronounced. Specifically, reducing hub vulnerability requires a careful balance between evacuation and dilution strategies. Evacuation involves decreasing $\kappa$ to promote outflow from the hub and increasing $\delta$ to limit inflow from the leaf. In contrast, dilution involves increasing $\kappa$ to retain the hub population and decreasing $\delta$ to encourage migration from the leaf to the hub. Minimizing hub vulnerability necessitates an optimal trade-off between these competing strategies.

Figure~\ref{fig:3}\textbf{b} illustrates that the optimal balance between evacuation and dilution strategies occurs along the downward diagonal, where $\kappa \approx 1 - \delta$. This result can also be derived from Eq.~\eqref{eq:equation2}, assuming $\beta \ll 1$. Deviations from this equilibrium, such as lower $\kappa$ and lower $\delta$, lead to increased exposure of the incoming leaf population in the hub due to the reduced size of the hub population. Conversely, higher $\kappa$ and higher $\delta$ result in the hub population being permanently exposed to the disease without adequate dilution from the leaf population. (Refer to Supplementary Materials Fig. S5 for other intermediate values in the range $0 \leq \alpha \leq 1$.)



\subsection{Non-pharmaceutical intervention (NPI) strategies to mitigate the vulnerability}
\begin{figure}[t]
    \centering
    \includegraphics[width=16cm]{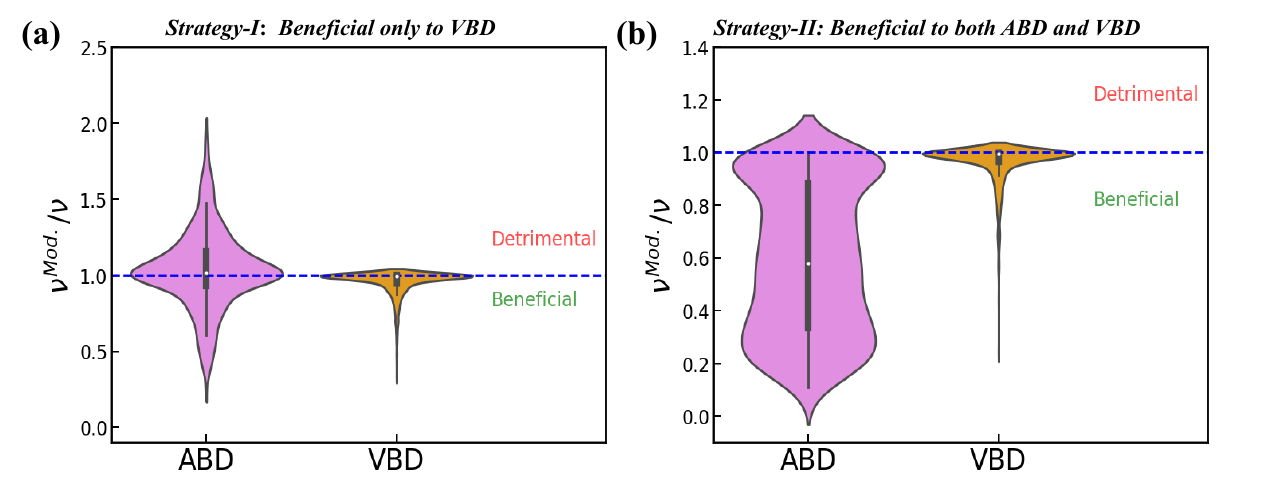}
    \caption{\textbf{NPI strategies to mitigate the epidemic vulnerability.} \textbf{(a)} Violin plot showing the impact of reshuffling strategy I, which adjusts mobility parameters within the constrained range $\delta = 1 - \kappa$, on epidemic vulnerability for ABDs and VBDs. \textbf{(b)} Violin plot illustrating the effect of reshuffling strategy II, where mobility parameters are set to specific values $\kappa = \frac{1}{\gamma + 1}$ and $\delta = \frac{\gamma}{\gamma + 1}$, on epidemic vulnerability for ABDs and VBDs. In both panels, the numerator ($\nu^{Mod}$) represents the recomputed epidemic vulnerability after reshuffling. The lower zone (values below 1) indicates a beneficial effect, while the upper zone (values above 1) signifies a detrimental effect. All parameters, except for those involved in the reshuffling, are sampled randomly from a uniform distribution.}
    \label{fig:4}
\end{figure}

In the absence of universally effective vaccines and therapeutics, non-pharmaceutical interventions (NPIs) play a critical role in controlling epidemics. These interventions often involve regulating travel behavior and mobility flows within populations. By studying the contagion dynamics of various diseases, we have uncovered important relationships between epidemic vulnerability and mobility patterns. While strict lockdowns are effective in curbing outbreaks, they come with significant socioeconomic costs. Previous analyses of Fig.~\ref{fig:figure1}\textbf{b} demonstrated that redistributing mobility---rather than enforcing complete lockdowns---by shifting flows from hotspots to neighboring suburban regions can aid in controlling ABDs in Cali, though it may exacerbate the spread of VBDs. Building on this understanding, this section examines the potential benefits of reshuffling mobility flows as an NPI strategy for managing both ABDs and VBDs.
Based on the derived vulnerability equations, ABD vulnerability, Eq.~\eqref{eq:equation1}, is minimized at specific mobility parameter values: $\kappa \approx \frac{1}{\gamma+1}$ and $\delta \approx \frac{\gamma}{\gamma+1}$. In contrast, VBD vulnerability, Eq.~\eqref{eq:equation2}, can be reduced over a broader range of mobility parameter values, including $\delta = 1 - \kappa$, irrespective of population size. These findings are consistent with the contagion dynamics analysis for both disease types across various values of $\alpha$. Notably, the optimal mobility parameter setting for ABDs is a subset of the broader $\delta = 1 - \kappa$ range identified for VBDs.

Next, we investigate two reshuffling strategies: (1) tuning mobility parameters to the specific values $\kappa = \frac{1}{\gamma+1}$ and $\delta = \frac{\gamma}{\gamma+1}$, and (2) adjusting mobility parameters within the constrained range $\delta = 1 - \kappa$, to determine whether either strategy is beneficial for both ABDs and VBDs.
To evaluate the effectiveness of these strategies, we generate multiple synthetic networks by exploring the parameter space defined by $(\alpha, \gamma, \beta, \delta, \kappa)$. For each network, we calculate the vulnerability $\nu$ before and after the intervention, denoted as $\nu^{Mod}$. By analyzing the resulting histogram of vulnerability ratios, we assess whether the reshuffling strategy is beneficial or detrimental based on the frequency of values less than or greater than 1.

\subsubsection{Strategy-I: Constraining mobility parameters to the range $\delta = 1-\kappa$}

Strategy I involves modifying the mobility flows from the leaf such that $\delta = 1 - \kappa$. Based on our earlier analysis, this intervention is anticipated to reduce vulnerability for VBDs; however, its effectiveness for ABDs is less certain. By performing multiple random iterations across synthetic networks, we observe that for VBDs, the majority of outcomes fall within the beneficial region of Fig.~\ref{fig:4}\textbf{a}, confirming the expected positive impact of this strategy. 
In contrast, the results for ABDs are more variable. Approximately half of the computed values lie within the beneficial region, while the remaining half fall in the detrimental region. This mixed outcome underscores the limitations of Strategy I for ABDs, as the precise tuning required to minimize their vulnerability is not consistently achieved. This discrepancy is likely due to the narrower optimal parameter range for ABDs ($\kappa \approx \frac{1}{\gamma + 1}$ and $\delta \approx \frac{\gamma}{\gamma + 1}$), which Strategy I does not explicitly satisfy.
These findings suggest that while Strategy I is a robust approach for reducing vulnerability in VBDs, its application to ABDs is less reliable and may require additional adjustments or complementary interventions to ensure its effectiveness.

\subsubsection{Strategy-II: Fixing mobility parameters to $\kappa=\frac{1}{\gamma+1}$ \& $\delta=\frac{\gamma}{\gamma+1}$}

Our analysis of contagion dynamics reveals that vulnerability to ABDs is minimized at specific mobility parameter values: $\kappa = \frac{1}{\gamma+1}$ and $\delta = \frac{\gamma}{\gamma+1}$. Notably, these optimal parameters for ABDs fall within the constrained range $\delta = 1 - \kappa$, where vulnerability to VBDs is also minimized. Consequently, setting these specific mobility parameters is expected to reduce vulnerability for both ABDs and VBDs. This expectation is confirmed by Fig.~\ref{fig:4}\textbf{(b)}, where the majority of vulnerability ratios ($\nu^{Mod}/\nu$) fall below 1, indicating that this intervention---referred to as Strategy II---is effective in reducing vulnerability for both disease types.

These findings highlight a direct relationship between vulnerability and the spatial scale (relative area) of the leaf compared to the hub, characterized by the parameter $\gamma$. Optimizing mobility based on this ratio emerges as a beneficial policy. Specifically, within-hub mobility ($\kappa$) should be adjusted inversely proportional to the relative size of the leaf. For instance, if the leaf area is significantly larger than the hub area ($\gamma \gg 1$), reducing within-hub mobility by migrating the majority of the hub population to the leaf ($\kappa \approx 0$) can decrease disease vulnerability. The larger area of the leaf disperses individuals, reducing contact rates and, consequently, exposure to airborne pathogens compared to the denser hub.
Additionally, implementing an outward lockdown in the leaf region ($\delta = 1$) can further minimize vulnerability by preventing the movement of infected individuals back to the hub. This approach ensures that individuals remain in the lower-density environment of the leaf, reducing opportunities for disease transmission. However, when the leaf and hub areas are comparable ($\gamma \approx 1$), complete migration of the hub population and a full outward restriction of the leaf population may not be optimal. In such cases, evenly distributing the population between both patches can yield the best outcome by optimizing the spread of the population and minimizing disease contact rates.

\subsection{Application of NPI Strategies to Cali}
\begin{figure}[t]
    \centering
    \includegraphics[width=16cm]{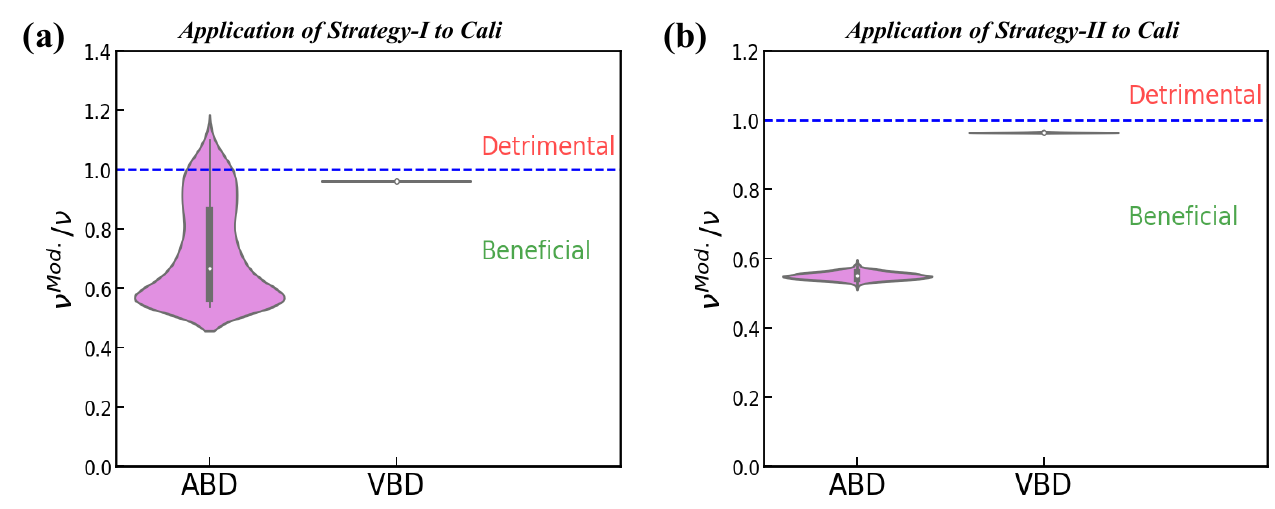}
    \caption{\textbf{Application of NPI strategies to to Cali, Colombia.}  
    Violin plots illustrating the impact of reshuffling flows using Strategy-I ({\bf a}) and Strategy-II ({\bf b}) on the epidemic vulnerability of Cali.
    }
    \label{fig:figure5}
\end{figure}

We now apply the strategies developed for the simplified one-hub-leaf model to the complex mobility network of Cali, demonstrating the practicality and effectiveness of our approach in real-world scenarios. As shown in Fig.~\ref{fig:figure1}\textbf{b}, a previously proposed NPI strategy---redistributing mobility between hotspots and suburbs while maintaining total flow volume---proved effective for mitigating ABDs but was detrimental for VBDs in Cali. In the synthetic model, this strategy corresponds to setting $\kappa = 0$.

In Fig.~\ref{fig:figure5}\textbf{a}, we evaluate the application of Strategy I (described in Section S8 of the Supplementary Material) to Cali's complex mobility network. While Strategy I was less effective for ABDs in the one-hub-leaf model, its application to Cali produced predominantly beneficial results, with only a small number of detrimental outcomes. For VBDs, the one-hub-leaf model showed mostly beneficial results, and the application to Cali yielded even more robust outcomes, with all results falling within the beneficial region.

We then applied Strategy II to the complex mobility network of Cali. Strategy II involves setting the mobility parameters $\kappa = \frac{1}{\gamma+1}$ and $\delta = \frac{\gamma}{\gamma+1}$. To implement this approach, we first calculated the value of $\gamma$ for Cali's network. Extending the definitions of $\alpha$, $\beta$, and $\gamma$ from the one-hub-leaf model to the complex network, we defined these parameters as the ratios of aggregated values in suburban and hotspot regions. Specifically, $\gamma$ was calculated as the ratio of the total suburban area to the total hotspot area. For Cali, $\gamma$ was determined to be 1.2.
The application of this strategy is shown in Fig.~\ref{fig:figure5}\textbf{b}, with detailed implementation provided in Section S8 of the Supplementary Material. The results demonstrate that, for both ABDs and VBDs, the vulnerability ratios ($\frac{\nu^{Mod}}{\nu}$) fall entirely within the beneficial zone. This outcome aligns closely with the results from the one-hub-leaf model, where the majority of vulnerability ratios were below 1, indicating a consistent reduction in vulnerability.

This agreement between theoretical predictions and real-world data suggests that, given the observed commuting patterns, spatial distribution of vectors, and population demographics in Cali, the proposed mitigation strategies can effectively minimize the impact of potential outbreaks of both airborne and vector-borne diseases. Notably, these results hold true not just for a specific set of mobility parameter values but across a wide range of parameter combinations, highlighting the robustness and generalizability of our findings.

\section {Discussion}
This study introduces a unified framework for analyzing the dynamics of airborne diseases (ABDs) and vector-borne diseases (VBDs), integrating theoretical modeling with real-world data to explore the interplay between population distribution, mobility patterns, and epidemic vulnerability. By examining both disease types, we demonstrated how non-pharmaceutical interventions (NPIs) can be effectively tailored to mitigate disease spread while accounting for their distinct transmission dynamics.

Our findings emphasize the importance of spatial heterogeneity and mobility dynamics in shaping vulnerability. The simplified one-hub-leaf metapopulation model identified specific mobility parameters ($\kappa = \frac{1}{\gamma+1}$ and $\delta = \frac{\gamma}{\gamma+1}$) as optimal for reducing ABD vulnerability, balancing population redistribution and localized exposure. For VBDs, a broader parameter range, including $\delta = 1 - \kappa$, was effective, allowing greater flexibility in intervention design. The overlap in these optimal ranges suggests the feasibility of unified strategies addressing both disease types.

Applying this framework to the complex mobility network of Santiago de Cali, Colombia, validated the model's findings. Strategy II, involving fine-tuned mobility adjustments, consistently reduced vulnerabilities for both ABDs and VBDs, proving its robustness and practicality as a non-pharmaceutical intervention. While Strategy I, which redistributes flows based on a specific constraint, also yielded beneficial outcomes, its effectiveness for ABDs was less consistent, reinforcing the need for disease-specific adaptations.

The implications for public health policy are significant. For ABDs, interventions should prioritize targeted mobility adjustments that promote population homogeneity and reduce susceptibility in hotspots. For VBDs, strategies must incorporate environmental and vector-related factors alongside mobility interventions. The framework's ability to distill complex mobility networks into hub-leaf structures offers a practical tool for identifying critical regions and optimizing resource allocation. Its adaptability to varying parameter combinations underscores its potential application in other urban settings with similar dynamics.

Nonetheless, the study has limitations. The one-hub-leaf model, while valuable for theoretical insights, simplifies urban networks and may omit important spatial and demographic complexities. Extending the framework to account for multiple hubs, hierarchical mobility patterns, and dynamic factors such as seasonal changes, socio-economic disruptions, and healthcare accessibility could enhance its applicability~\cite{bassolas2019hierarchical, poudyal2024dynamic}. Incorporating additional data, such as vector ecology (e.g., breeding site distributions)~\cite{muir1998aedes}, could further refine vulnerability assessments, particularly for VBDs.
Focusing on Cali provided a strong validation of the model but limits its generalizability. Future research should apply this framework across diverse urban contexts to assess its broader applicability and improve its predictive power. Expanding data integration, including socio-economic disparities, healthcare infrastructure, and cultural mobility patterns, could further enhance the framework's real-world utility.

This work bridges theoretical modeling and practical application, offering an evidence-based approach for designing NPIs. By addressing the distinct dynamics of ABDs and VBDs, it lays a foundation for more targeted, effective mitigation strategies. Collaborative efforts among researchers, public health officials, and policymakers are essential to ensure interventions are equitable, context-sensitive, and impactful. While this study advances understanding, addressing the multifaceted challenges of epidemic vulnerability requires continued research and coordinated public health efforts to develop comprehensive and adaptive responses.

\section*{Acknowledgements}
GG acknowledges partial support through Grant No. 62417 from the John Templeton Foundation. The opinions expressed in this publication are those of the author(s) and do not necessarily reflect the views of the Foundation. GG and BP also acknowledge support from the University of Rochester. D.S-P acknowledges financial support through grants JDC2022-048339-I and PID2021-128005NB- C21 funded by MCIN/AEI/10.13039/501100011033 and the European Union “NextGenerationEU”/PRTR”.

\bibliography{ms}

\clearpage

\end{document}


\makeatletter
\renewcommand{\maketitle}{\bgroup\setlength{\parindent}{0pt}
	\begin{flushleft}
{\huge Supplementary Information}\\[0.3cm]
		{\large\textbf \@title}
		
		\@author
		
	\end{flushleft}\egroup
}
\makeatother
\title{Contrasting and comparing the efficacy of non-pharmaceutical interventions on air-borne and vector-borne diseases}

\newcommand{\floor}[1]{\lfloor #1 \rfloor}

\author{Bibandhan \surname{Poudyal}, David  \surname{Soriano Pan\~os}, and Gourab \surname{Ghoshal}}                         

\date{}

\maketitle

\tableofcontents
\newpage
\section{Mobility Matrix}
\begin{figure}[H]
    \centering
    \includegraphics[width=16cm]{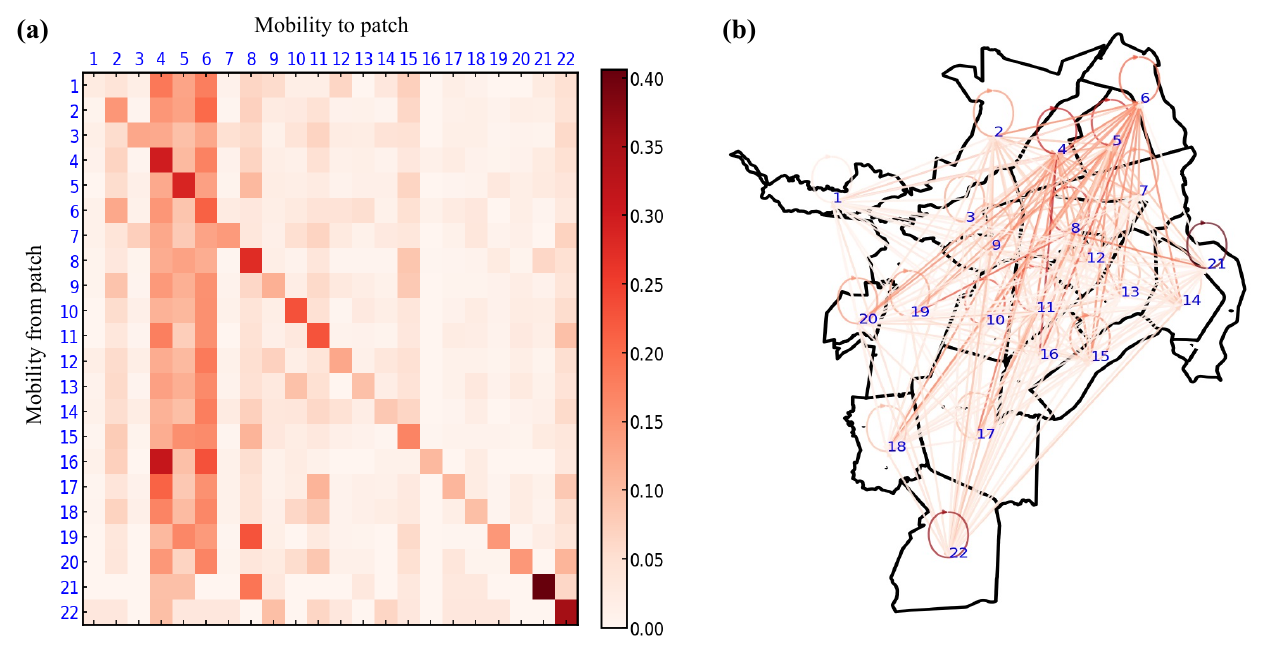}
    \caption{\textbf{Mobility in Cali.} \textbf{(a)} Weighted mobility matrix (\textbf{R}) displaying the normalized flow of individuals between patches. The matrix is normalized such that the total outgoing flows from each patch sum to 1. \textbf{(b)} Map of Cali illustrating the spatial distribution of patches alongside the associated mobility flows.}
    \label{fig:S1}
\end{figure}
\newpage
\section{Hotspot Classification}
\begin{figure}[H]
    \centering
    \includegraphics[width=16cm]{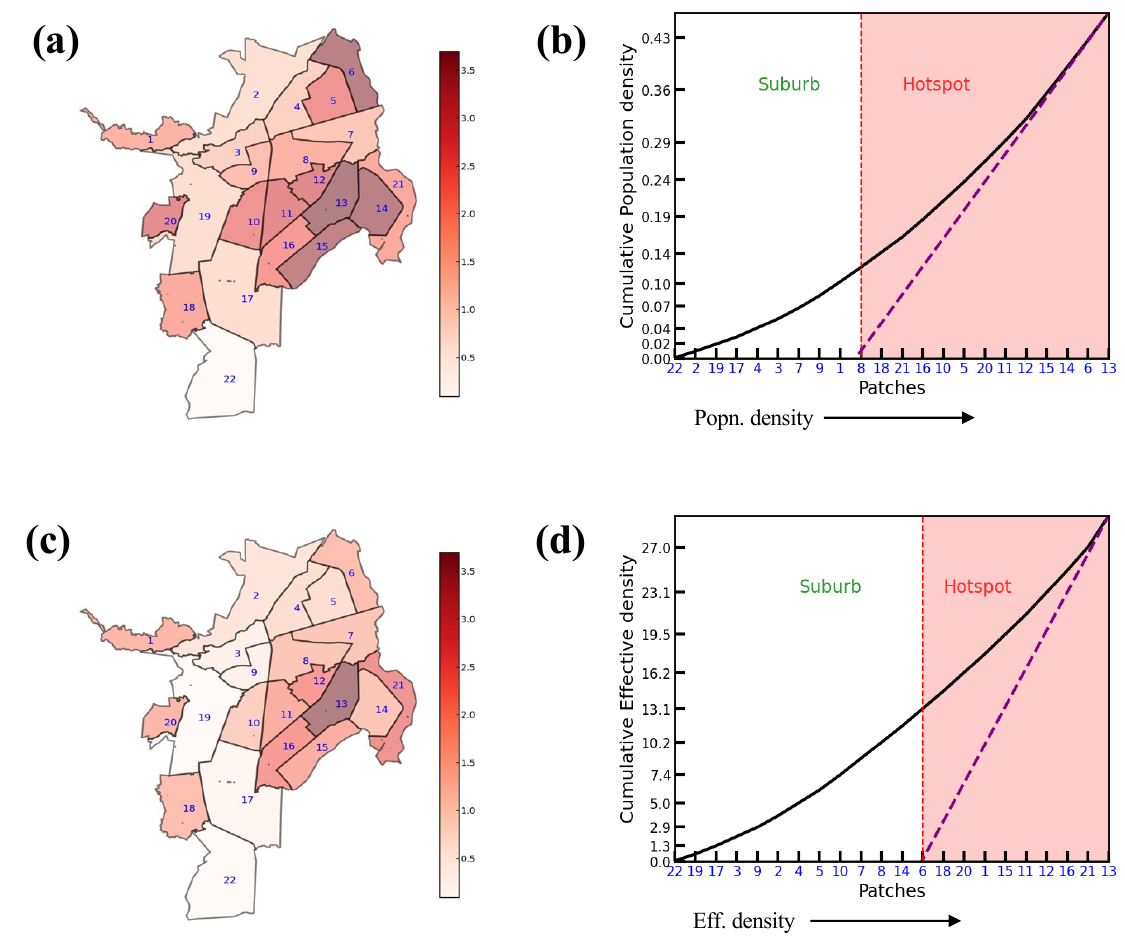}
    \caption{\textbf{Hotspot Classification using LouBar Method in Cali.} \textbf{(a)} Map of Cali depicting the spatial distribution of population density (in units of $10^{-4},\text{km}^{-2}$). \textbf{(b)} Lorenz curve representing the distribution of population density across patches, with the x-axis showing patches arranged in ascending order of population density and the y-axis indicating the cumulative population density. The intersection of the tangent line with the x-axis determines the threshold for hotspot classification. \textbf{(c)} Map of Cali illustrating the spatial distribution of effective density (mosquito-to-human ratio). \textbf{(d)} Lorenz curve showing the distribution of effective density across patches. The Spearman rank correlation coefficient between population density and effective density is 0.74, reflecting a strong positive correlation and indicating significant overlap between hotspots identified using these two metrics.}
    \label{fig:S2}
\end{figure}
\newpage
\section{Critical Matrices}
\subsection{ABD}
Assuming a negligible initial infectious seed size ($\epsilon \ll 1$) and no individuals in the recovery phase initially, the contagion dynamics of ABD can be described as follows [42,55]: \begin{align} 
\frac{\mu}{\lambda}\epsilon_i=\sum_{j=1}^{N}\underbrace{\Bigg[(1-p)^2\delta_{ij}\frac{f_i}{n^{eff}_i}+p(1-p)\Bigg(R_{ij}\frac{f_j}{n^{eff}_j}+R_{ji}\frac{f_i}{n^{eff}_i}\Bigg)+p^2\sum_lR_{il}R_{jl}\frac{f_l}{n^{eff}_l}\Bigg]n_j}_{M_{ij}}\epsilon_j, 
\end{align} where $\mu$ represents the recovery rate, $\lambda$ the contagion rate, $\epsilon_i$ the infectious size in patch $i$, and $p$ the probability that a healthy individual decides to move. Additionally, $f_i$ is the number of daily contacts per individual in patch $i$, $n^{eff}_i$ is the effective population of patch $i$ (accounting for individuals present but not residing in the location), $n_j$ is the human population in patch $j$, and \textbf{M} is the critical matrix, with $\frac{\mu}{\lambda}$ as its eigenvalue.

Assuming $\mu$ remains constant, the maximum eigenvalue of \textbf{M} determines the minimum contagion rate required for an epidemic outbreak. This maximum eigenvalue is defined as the epidemic vulnerability. The analysis prioritizes collective mobility patterns across patches rather than individual-scale movements. Therefore, without loss of generality, we assume that the probability of a healthy individual undertaking a trip, $p$, equals 1. This assumption does not preclude individuals from remaining within their local patch, as such behavior can be modeled as a zero-distance trip. Instead, it ensures that the framework captures all possible local mobility patterns within the defined geographic area. The critical matrix (\textbf{C} = \textbf{M}) under this assumption simplifies to:
\begin{align}
    C_{ij}^{ABD} = n_j \sum_{l=1}^{N} R_{il}R_{jl} \frac{f_l}{n_{eff}^l}.
\end{align}
Assuming that the number of contacts that each individual makes inside each patch depends on its density, we can substitute $f_l \approx \frac{n_{eff}^l}{a_l}$ where $a_l$ is the area of the patch l. Thus,
\begin{align}
    C_{ij}^{ABD} = n_j \sum_{l=1}^{N} \frac{R_{il}R_{jl}}{a_l}.
\end{align}
This gives:
\begin{align}
    C_{ii}^{ABD} &= n_i\Big(\frac{R_{ii}^2}{a_i} + \frac{R_{ij}^2}{a_j}\Big),\\
    C_{ij}^{ABD} &= n_j\Big(\frac{R_{ii}R_{ji}}{a_i} + \frac{R_{ij}R_{jj}}{a_j}\Big),\\
    C_{ji}^{ABD} &= n_i\Big(\frac{R_{ji}R_{ii}}{a_i} + \frac{R_{jj}R_{ij}}{a_j}\Big),\\
    C_{jj}^{ABD} &= n_j\Big(\frac{R_{ji}^2}{a_i} + \frac{R_{jj}^2}{a_j}\Big).
\end{align}
Hence,
\begin{align} \label{CM_ABD}
    C^{ABD} = \frac{n_i}{a_i}\Bigg(\begin{pmatrix}
R_{ii}^2 & \frac{n_j}{n_i}R_{ii}R_{ji}\\
R_{ii}R_{ji} & \frac{n_j}{n_i}R_{ji}^2
\end{pmatrix} + \frac{a_i}{a_j}\begin{pmatrix}
R_{ij}^2 & \frac{n_j}{n_i}R_{ij}R_{jj}\\
R_{ij}R_{jj} & \frac{n_j}{n_i}R_{jj}^2
\end{pmatrix}\Bigg).
\end{align}

\newpage
\subsection{VBD}
Similarly, assuming a negligible initial human and vector infectious seed size ($\epsilon^H<<1$ and $\epsilon^M<<1$) and no initial individuals in the recovery phase, the contagion dynamics of ABD can be described as follows [45]:
\begin{align}
    \epsilon^H_i = \sum_{j=1}^{N}\frac{\lambda^{MH}\beta^*}{\mu_H}\underbrace{\Bigg(pR_{ij}\frac{m_j}{n^{eff}_j} + (1-p)\delta_{ij}\frac{m_i}{n^{eff}_i}\Bigg)}_{M_{ij}}\epsilon^M_j,
\end{align}
\begin{align}
    \epsilon^M_i=\sum_{j=1}^{N}\frac{\lambda^{HM}\beta^*}{\mu_M}\underbrace{\Bigg(\alpha^*pR_{ij}\frac{n_j}{n^{eff}_i}+(1-\alpha^*p)\delta_{ij}\frac{n_i}{n^{eff}_i}\Bigg)}_{\Tilde{M}_{ij}}\epsilon^H_j,
\end{align}
where $\epsilon^{H_i}$ is the infectious size of humans in patch $i$, $\lambda^{MH}$ is the contagion rate of an infected vector transmitting the disease to a healthy individual, $\beta^{\ast}$ is the feeding rate of vectors, $\mu_H$ is the recovery rate of humans, $p$ is the probability that a healthy individual decides to move, $m_j$ is the vector population in patch $j$, and $n^{eff}_j$ is the effective population of location $j$, representing the number of humans present in the location but not residing there. 

Additionally, $\lambda^{HM}$ is the contagion rate of an infected human transmitting the disease to a healthy vector, $\mu_M$ is the mortality rate of vectors, $\alpha^{\ast}$ is a scaling factor that reflects the degree to which an ill individual can move (with $\alpha^{\ast} \in [0,1]$, depending on the nature of the disease and individuals' coping ability), and $n_j$ is the human population in patch $j$.
It is important to note that the parameters $\beta^{\ast}$ and $\alpha^{\ast}$ in the above equations are distinct from the leaf-to-hub ratios of human population ($\alpha$) and vector population ($\beta$) used elsewhere in this paper. The parameters $\beta^{\ast}$ and $\alpha^{\ast}$ specifically describe behavioral and biological dynamics within the vector-host interaction framework.

Thus, the bipartite nature of VBD transmission, involving vector-to-human and human-to-vector infections, can be represented respectively using matrices $M$ and $\Tilde{M}$ as:
\begin{align}
    M_{ij} &= pR_{ij}\frac{m_j}{n^{eff}_j}+(1-p)\delta_{ij}\frac{m_i}{n^{eff}_i},\\
    \Tilde{M}_{ij} &= \alpha^*pR_{ji}\frac{n_j}{n^{eff}_i}+(1-\alpha^*p)\delta_{ij}\frac{n_i}{n^{eff}_i}.
\end{align}
The nontrivial solutions for $\epsilon_H$ correspond to the eigenvectors of matrix $M\Tilde{M}$. Thus, the critical matrix defining VBD epidemic vulnerability can be defined as: $C = M\Tilde{M}$. As we are more interested towards comprehensive patch mobility rather than mobility in individual scale; without loss of generality we assume $p=1$ \& $\alpha^*=1$. The critical matrix (C) thus becomes:
\begin{align}
    C_{ij}^{VBD} = n_j \sum_{l=1}^{N} R_{il}R_{jl} \frac{m_l}{(n_{eff}^l)^2}.
\end{align}
Here, the effective population $n_{eff}^i$ is defined as:
\begin{align}
    n_{eff}^i = \sum_{j=i}n_jR_{ji}.
\end{align}

Thus, for two patch i \& j, the above equation can be written as:
\begin{align}
    n_{eff}^i = n_iR_{ii}+n_jR_{ji},\\
    n_{eff}^j = n_iR_{ij}+n_jR_{jj}.
\end{align}
Similarly the critical matrix can be written as:
\begin{align}
    C_{ii}^{VBD} &= \frac{m_i}{n_i}\left(\frac{R_{ii}^2}{(R_{ii} + \frac{n_j}{n_i} R_{jj})^2} + \frac{\frac{m_j}{m_i}R_{ij}^2}{(R_{ij} + \frac{n_j}{n_i} R_{jj})^2}\right),\\
    C_{ij}^{VBD} &= \frac{m_i}{n_i}\left(\frac{\frac{n_j}{n_i} R_{ii}R_{ji}}{(R_{ii} + \frac{n_j}{n_i}  R_{jj})^2} + \frac{\frac{n_j}{n_i}\frac{m_j}{m_i}R_{ij}R_{jj}}{(R_{ij} + \frac{n_j}{n_i} R_{jj})^2}\right),\\
    C_{ji}^{VBD} &= \frac{m_i}{n_i}\left(\frac{R_{ii}R_{ji}}{(R_{ii} + \frac{n_j}{n_i} R_{jj})^2}+ \frac{\frac{m_j}{m_i}R_{ij}R_{jj}}{(R_{ij} + \frac{n_j}{n_i} R_{jj})^2}\right),\\
    C_{jj}^{VBD} &= \frac{m_i}{n_i}\left(\frac{\frac{n_j}{n_i}R_{ji}^2}{(R_{ii} + \frac{n_j}{n_i} R_{jj})^2} + \frac{\frac{n_j}{n_i}\frac{m_j}{m_i} R_{jj}^2}{(R_{ij} + \frac{n_j}{n_i} R_{jj})^2}\right).\\
\end{align}
Hence,
\begin{align}\label{CM_VBD}
   C^{VBD}  = \frac{m_i}{n_i}\Bigg(\frac{1}{\Big(R_{ii}+\frac{n_j}{n_i}R_{ji}\Big)^2}\begin{pmatrix}
R_{ii}^2 & \frac{n_j}{n_i} R_{ii}R_{ji}\\
R_{ii}R_{ji} & \frac{n_j}{n_i}R_{ji}^2
\end{pmatrix} + \frac{\frac{m_j}{m_i}}{\Big(R_{ij}+\frac{n_j}{n_i} R_{jj})^2}\begin{pmatrix}
R_{ij}^2 & \frac{n_j}{n_i}R_{ij}R_{jj}\\
R_{ij}R_{jj} & \frac{n_j}{n_i} R_{jj}^2
\end{pmatrix}\Bigg).
\end{align}

\newpage
\section{Calculation of Epidemic Vulnerability}
 \begin{figure}[H]
    \centering
    \includegraphics[width=0.6\columnwidth]{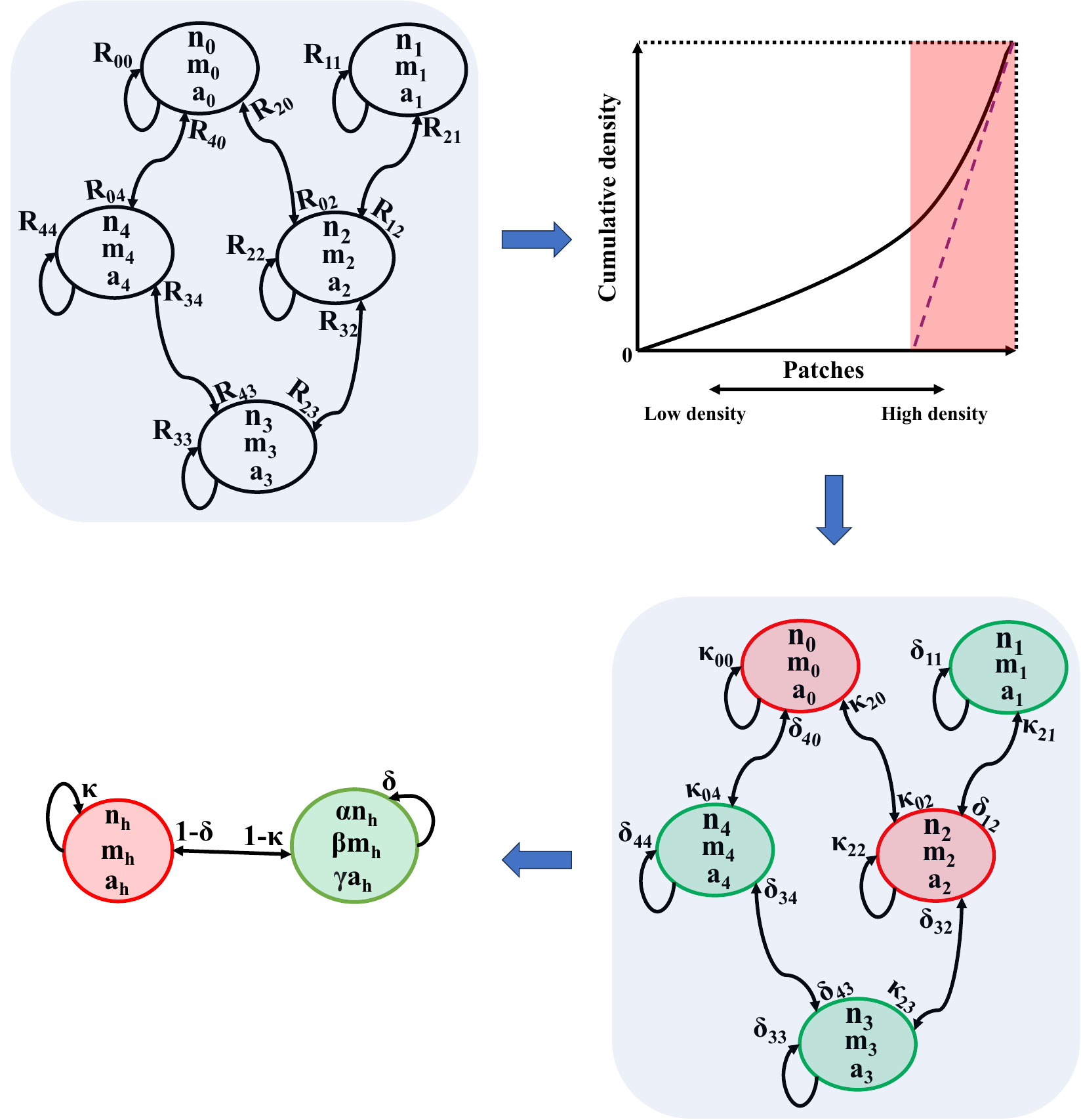}
    \caption{\textbf{Simplifying the Complex Model to a One Hub-Leaf Model.} Using the human population ($n$), vector population ($m$), and area of each patch ($a$), we calculate the population density for each patch. Hotspots are identified based on their densities using the LouBar method, with hotspots and suburbs (non-hotspots) represented by red and green patches, respectively. The mobility associated with hotspots is denoted by the parameter $\kappa$, while mobility in suburbs is denoted by $\delta$.
The complex model is then simplified into a one hub-leaf model, where the hub represents the cumulative hotspot and the leaf represents the cumulative non-hotspot (suburb). Overall mobility within the hub is represented by $\kappa$, while mobility within the leaf is represented by $\delta$. The model also incorporates the proportions of key variables between the hub and the leaf: $\alpha$ denotes the proportion of the human population in the leaf relative to the hub, $\beta$ denotes the proportion of the vector population in the leaf relative to the hub, and $\gamma$ denotes the proportion of the leaf area relative to the hub.}
    \label{fig:S3}
\end{figure}
\newpage
For one hub-leaf model, we define vulnerability of the hub (i.e. assuming $i$ to be hub and $j$ to be leaf) where human population $n_i=n_h$, vector population $m_i = m_h$, area $a_i = a_h$, $R_{ii}=\kappa$, $R_{ij}=1-\kappa$, $R_{jj}=\delta$, $R_{ji}=1-\delta$. We assume $\alpha$ denotes the proportion of human population in leaf  to the hub, $\beta$ denotes the proportion of vector population in leaf to the hub (these $\alpha$, $\beta$ is different than $\alpha^*$, $\beta^*$ above) and $\gamma$ denotes proportion of leaf area to the hub. Thus, the leaf human population $n_l=\alpha n_h$, leaf vector population $m_l=\beta m_h$ \& leaf area $a_l = \gamma a_h$.\\
\subsection{ABD}
The Critical Matrix of ABD defined in Eq. \ref{CM_ABD} becomes:
\begin{align}
   \implies C^{ABD} = \frac{n_h}{a_h}\Biggr[\begin{pmatrix}\kappa^2 & \alpha\kappa(1-\delta)\\
\kappa(1-\delta) & \alpha(1-\delta)^2
\end{pmatrix} + \frac{1}{\gamma}\begin{pmatrix}
(1-\kappa)^2 & \alpha(1-\kappa)\delta\\
(1-\kappa)\delta & \alpha\delta^2
\end{pmatrix}\Biggr]. \nonumber
\end{align}
\begin{align}
\implies   C^{ABD} = \frac{n_h}{a_h}\begin{pmatrix}\kappa^2+\frac{(1-\kappa)^2}{\gamma} & \alpha\kappa(1-\delta)+\frac{\alpha(1-\kappa)\delta}{\gamma}\\
\kappa(1-\delta)+\frac{(1-\kappa)\delta}{\gamma} & \alpha(1-\delta)^2+\frac{\alpha\delta^2}{\gamma}.
\end{pmatrix} 
\end{align}
The scale factor $\frac{n_h}{a_h}$ does not influence the contagion dynamics and can therefore be neglected. Thus, the eigenvalue for $C^{ABD}$ is:
\begin{align}
    |C^{ABD}-\lambda I| = 0 \nonumber
\end{align}
\begin{align}
    \implies \begin{vmatrix} \kappa^2+\frac{(1-\kappa)^2}{\gamma}-\lambda & \alpha\kappa(1-\delta)+\frac{\alpha(1-\kappa)\delta}{\gamma}\\
\kappa(1-\delta)+\frac{(1-\kappa)\delta}{\gamma} & \alpha(1-\delta)^2+\frac{\alpha\delta^2}{\gamma}-\lambda
\end{vmatrix}  = 0 \nonumber
\end{align}
Let's say $X = \kappa(1-\delta)+\frac{\alpha(1-\kappa)\delta}{\gamma}$,
\begin{align}
    \implies \begin{vmatrix} \kappa^2+\frac{(1-\kappa)^2}{\gamma}-\lambda & \alpha X\\
X & \alpha(1-\delta)^2+\frac{\alpha\delta^2}{\gamma}-\lambda
\end{vmatrix}  = 0. \nonumber
\end{align}
\begin{align*}
    \implies &\underbrace{\alpha\kappa^2(1-\delta)^2} + \frac{\alpha\kappa^2\delta^2}{\gamma} + \frac{\alpha(1-\kappa)^2(1-\delta)^2}{\gamma} + \underbrace{\frac{\alpha\delta^2(1-\kappa)^2}{\gamma^2}} - \Big(\kappa^2+\alpha(1-\delta)^2 + \frac{(1-\kappa)^2+\alpha\delta^2}{\gamma}\Big) \lambda \\
    &+ \lambda^2 -\alpha X^2= 0. \nonumber
\end{align*}
\begin{align*}
    \implies &\underbrace{\alpha\kappa^2(1-\delta)^2+\frac{\alpha\delta^2(1-\kappa)^2}{\gamma^2}+\textcolor{red}{\frac{2\alpha\kappa\delta(1-\kappa)(1-\delta)}{\gamma}}}-\textcolor{red}{\frac{2\alpha\kappa\delta(1-\kappa)(1-\delta)}{\gamma}} + \frac{\alpha\kappa^2\delta^2}{\gamma}\\
    & + \frac{\alpha(1-\kappa)^2(1-\delta)^2}{\gamma}
    - \Big(\kappa^2+\alpha(1-\delta)^2 + \frac{(1-\kappa)^2+\alpha\delta^2}{\gamma}\Big) \lambda + \lambda^2 -\alpha X^2= 0. \nonumber
\end{align*}
\begin{align*}
    \implies &\cancel{\alpha X^2}-\underbrace{\frac{2\alpha\kappa\delta(1-\kappa)(1-\delta)}{\gamma} + \frac{\alpha\kappa^2\delta^2}{\gamma} + \frac{\alpha(1-\kappa)^2(1-\delta)^2}{\gamma}} - \Big(\kappa^2+\alpha(1-\delta)^2 + \frac{(1-\kappa)^2+\alpha\delta^2}{\gamma}\Big) \lambda \\
    & + \lambda^2 -\cancel{\alpha X^2}= 0. \nonumber
\end{align*}
\begin{align}
    \implies \frac{\alpha}{\gamma}\underbrace{\Big(\kappa\delta-(1-\kappa)(1-\delta)\Big)^2} - \Big(\kappa^2+\alpha(1-\delta)^2 + \frac{(1-\kappa)^2+\alpha\delta^2}{\gamma}\Big) \lambda + \lambda^2= 0. \nonumber
\end{align}
\begin{align}
    \implies \frac{\alpha}{\gamma}(1-\kappa-\delta)^2 - \Big(\kappa^2+\alpha(1-\delta)^2 + \frac{(1-\kappa)^2+\alpha\delta^2}{\gamma}\Big) \lambda + \lambda^2= 0. \nonumber
\end{align}
Which is in the form of $ax^2+bx+c=0$ with the solution:
\begin{align}
    \lambda = \frac{\kappa^2+\alpha(1-\delta)^2 + \frac{(1-\kappa)^2+\alpha\delta^2}{\gamma}\pm \sqrt{\Big(\kappa^2+\alpha(1-\delta)^2 + \frac{(1-\kappa)^2+\alpha\delta^2}{\gamma}\Big)^2 - 4\frac{\alpha}{\gamma}(1-\kappa-\delta)^2}}{2}.
\end{align}
Thus,
\begin{align}
    \lambda_{ABD}^{max} = \frac{\kappa^2+\alpha(1-\delta)^2 + \frac{(1-\kappa)^2+\alpha\delta^2}{\gamma} + \sqrt{\Big(\kappa^2+\alpha(1-\delta)^2 + \frac{(1-\kappa)^2+\alpha\delta^2}{\gamma}\Big)^2 - 4\frac{\alpha}{\gamma}(1-\kappa-\delta)^2}}{2}.
\end{align}
We define Vulnerability in ABD ($\nu^{ABD}$) as $\lambda_{ABD}^{max}$. Thus the expression for overall vulnerability of the system is:
\begin{align}\label{ABD}
    \boxed{\nu^{ABD} = \frac{\Theta_h(\kappa,\delta) + \frac{\Theta_l(\kappa,\delta)}{\gamma} + \sqrt{\Big(\Theta_h(\kappa,\delta) + \frac{\Theta_l(\kappa,\delta)}{\gamma} \Big)^2 - \frac{\alpha}{\gamma}\Big(\Theta_{hl}(\kappa,\delta)\Big)^2}}{2},}
\end{align}
where,
\begin{itemize}
    \item hub-vulnerability: $\Theta_h(\kappa,\delta)=\kappa^2+\alpha(1-\delta)^2$,
    \item leaf-vulnerability: $\Theta_l(\kappa,\delta)=(1-\kappa)^2+\alpha\delta^2$,
    \item cross-vulnerability: $\Theta_{hl}(\kappa,\delta)=2(1-\kappa-\delta)$.
\end{itemize}
\newpage
\subsection{VBD}
The Critical Matrix of VBD defined in Eq. \ref{CM_VBD} becomes:

\begin{align*}
 \implies   C^{VBD} = \frac{m_h}{n_h}\Biggr[&\frac{1}{\Big(\kappa+\alpha(1-\delta)\Big)^2}\begin{pmatrix}
\kappa^2 & \alpha\kappa(1-\delta)\\
\kappa(1-\delta) & \alpha(1-\delta)^2
\end{pmatrix} \\
&+ \frac{\beta}{\Big((1-\kappa)+\alpha\delta)^2}\begin{pmatrix}
(1-\kappa)^2 & \alpha(1-\kappa)\delta\\
(1-\kappa)\delta & \alpha\delta^2 
\end{pmatrix} \Biggr]. \nonumber
\end{align*}
\begin{align}
 \implies   C^{VBD}  = \frac{m_h}{n_h} \begin{pmatrix}
\frac{\kappa^2}{\Big(\kappa+\alpha(1-\delta)\Big)^2} +\frac{\beta(1-\kappa)^2}{\Big((1-\kappa)+\alpha\delta\Big)^2} & \frac{\alpha\kappa(1-\delta)}{\Big(\kappa+\alpha(1-\delta)\Big)^2} + \frac{\beta\alpha(1-\kappa)\delta}{\Big((1-\kappa)+\alpha\delta\Big)^2}\\
\frac{\kappa(1-\delta)}{\Big(\kappa+\alpha(1-\delta)\Big)^2} + \frac{\beta(1-\kappa)\delta}{\Big((1-\kappa)+\alpha\delta\Big)^2}& \frac{\alpha(1-\delta)^2}{\Big(\kappa+\alpha(1-\delta)\Big)^2} + \frac{\beta\alpha\delta^2}{\Big((1-\kappa)+\alpha\delta\Big)^2}
\end{pmatrix}.
\end{align}
The scale factor $\frac{m_h}{n_h}$ does not influence the contagion dynamics and can therefore be neglected. Thus, the eigenvalue for $C^{VBD}$ is:
\begin{align}
    |C^{VBD}-\lambda I| = 0 \nonumber
\end{align}
\begin{align}
 \implies   \begin{vmatrix}
\frac{\kappa^2}{\Big(\kappa+\alpha(1-\delta)\Big)^2} +\frac{\beta(1-\kappa)^2}{\Big((1-\kappa)+\alpha\delta\Big)^2}-\lambda & \frac{\alpha\kappa(1-\delta)}{\Big(\kappa+\alpha(1-\delta)\Big)^2} + \frac{\beta\alpha(1-\kappa)\delta}{\Big((1-\kappa)+\alpha\delta\Big)^2}\\
\frac{\kappa(1-\delta)}{\Big(\kappa+\alpha(1-\delta)\Big)^2} + \frac{\beta(1-\kappa)\delta}{\Big((1-\kappa)+\alpha\delta\Big)^2}& \frac{\alpha(1-\delta)^2}{\Big(\kappa+\alpha(1-\delta)\Big)^2} + \frac{\beta\alpha\delta^2}{\Big((1-\kappa)+\alpha\delta\Big)^2}-\lambda
\end{vmatrix} = 0. \nonumber
\end{align}
Let's say $X = \frac{\kappa(1-\delta)}{\Big(\kappa+\alpha(1-\delta)\Big)^2} + \frac{\beta(1-\kappa)\delta}{\Big((1-\kappa)+\alpha\delta\Big)^2}$,
\begin{align}
 \implies   \begin{vmatrix}
\frac{\kappa^2}{\Big(\kappa+\alpha(1-\delta)\Big)^2} +\frac{\beta(1-\kappa)^2}{\Big((1-\kappa)+\alpha\delta\Big)^2}-\lambda & \alpha X\\
X & \frac{\alpha(1-\delta)^2}{\Big(\kappa+\alpha(1-\delta)\Big)^2} + \frac{\beta\alpha\delta^2}{\Big((1-\kappa)+\alpha\delta\Big)^2}-\lambda
\end{vmatrix} = 0. \nonumber
\end{align}
\begin{align*}
  \implies  \underbrace{\frac{\kappa^2\alpha(1-\delta)^2}{\Big(\kappa+\alpha(1-\delta)\Big)^4}} + \frac{\kappa^2\beta\alpha\delta^2}{\Big(\kappa+\alpha(1-\delta)\Big)^2\Big((1-\kappa)+\alpha\delta\Big)^2}+\frac{\beta(1-\kappa)^2\alpha(1-\delta)^2}{\Big((1-\kappa)+\alpha\delta\Big)^2\Big(\kappa+\alpha(1-\delta)\Big)^2}\\
  +\underbrace{\frac{\beta^2\alpha(1-\kappa)^2\delta^2}{\Big((1-\kappa)+\alpha\delta\Big)^4}}
    -\Bigg[\frac{\kappa^2+\alpha(1-\delta)^2}{\Big(\kappa+\alpha(1-\delta)\Big)^2} + \frac{\beta\Big((1-\kappa)^2+\alpha\delta^2\Big)}{\Big((1-\kappa)+\alpha\delta\Big)^2}\Bigg] \lambda + \lambda^2 -\alpha X^2= 0.
\end{align*}
\begin{align*}
 \implies &\underbrace{\frac{\kappa^2\alpha(1-\delta)^2}{\Big(\kappa+\alpha(1-\delta)\Big)^4}+\frac{\beta^2\alpha(1-\kappa)^2\delta^2}{\Big((1-\kappa)+\alpha\delta\Big)^4}+\textcolor{red}{\frac{2\kappa\alpha(1-\delta)\beta(1-\kappa)\delta}{\Big(\kappa+\alpha(1-\delta)\Big)^2\Big((1-\kappa)+\alpha\delta\Big)^2}}}\\
 &- \textcolor{red}{\frac{2\kappa\alpha(1-\delta)\beta(1-\kappa)\delta}{\Big(\kappa+\alpha(1-\delta)\Big)^2\Big((1-\kappa)+\alpha\delta\Big)^2}}+\frac{\kappa^2\beta\alpha\delta^2}{\Big(\kappa+\alpha(1-\delta)\Big)^2\Big((1-\kappa)+\alpha\delta\Big)^2}\\
 &+\frac{\beta(1-\kappa)^2\alpha(1-\delta)^2}{\Big((1-\kappa)+\alpha\delta\Big)^2\Big(\kappa+\alpha(1-\delta)\Big)^2}-\Bigg[\frac{\kappa^2+\alpha(1-\delta)^2}{\Big(\kappa+\alpha(1-\delta)\Big)^2} + \frac{\beta\Big((1-\kappa)^2+\alpha\delta^2\Big)}{\Big((1-\kappa)+\alpha\delta\Big)^2}\Bigg] \lambda\\
 &+ \lambda^2 -\alpha X^2= 0.
\end{align*}
\begin{align*}
 \implies &\cancel{\alpha X^2}-\underbrace{\frac{2\kappa\alpha(1-\delta)\beta(1-\kappa)\delta}{\Big(\kappa+\alpha(1-\delta)\Big)^2\Big((1-\kappa)+\alpha\delta\Big)^2} +\frac{\kappa^2\beta\alpha\delta^2}{\Big(\kappa+\alpha(1-\delta)\Big)^2\Big((1-\kappa)+\alpha\delta\Big)^2}}\\
 &\underbrace{+\frac{\beta(1-\kappa)^2\alpha(1-\delta)^2}{\Big((1-\kappa)+\alpha\delta\Big)^2\Big(\kappa+\alpha(1-\delta)\Big)^2}}-\Bigg[\frac{\kappa^2+\alpha(1-\delta)^2}{\Big(\kappa+\alpha(1-\delta)\Big)^2} + \frac{\beta\Big((1-\kappa)^2+\alpha\delta^2\Big)}{\Big((1-\kappa)+\alpha\delta\Big)^2}\Bigg] \lambda \\
 &+ \lambda^2 -\cancel{\alpha X^2}= 0.
\end{align*}
\begin{align*}
 \implies \frac{\alpha\beta\underbrace{\Big(\kappa\delta-(1-\kappa)(1-\delta)\Big)^2}}{\Big(\kappa+\alpha(1-\delta)\Big)^2\Big((1-\kappa)+\alpha\delta\Big)^2}-\Bigg[\frac{\kappa^2+\alpha(1-\delta)^2}{\Big(\kappa+\alpha(1-\delta)\Big)^2} + \frac{\beta\Big((1-\kappa)^2+\alpha\delta^2\Big)}{\Big((1-\kappa)+\alpha\delta\Big)^2}\Bigg] \lambda + \lambda^2 = 0.
\end{align*}
\begin{align*}
 \implies \frac{\alpha\beta(1-\kappa-\delta)^2}{\Big(\kappa+\alpha(1-\delta)\Big)^2\Big((1-\kappa)+\alpha\delta\Big)^2}-\Bigg[\frac{\kappa^2+\alpha(1-\delta)^2}{\Big(\kappa+\alpha(1-\delta)\Big)^2} + \frac{\beta\Big((1-\kappa)^2+\alpha\delta^2\Big)}{\Big((1-\kappa)+\alpha\delta\Big)^2}\Bigg] \lambda + \lambda^2 = 0.
\end{align*}
Which is in the form of $ax^2+bx+c=0$ with the solution:
\begin{align}
    \lambda = \frac{\frac{\kappa^2+\alpha(1-\delta)^2}{\Big(\kappa+\alpha(1-\delta)\Big)^2} + \frac{\beta\Big((1-\kappa)^2+\alpha\delta^2\Big)}{\Big((1-\kappa)+\alpha\delta\Big)^2}\pm \sqrt{\Bigg[\frac{\kappa^2+\alpha(1-\delta)^2}{\Big(\kappa+\alpha(1-\delta)\Big)^2} + \frac{\beta\Big((1-\kappa)^2+\alpha\delta^2\Big)}{\Big((1-\kappa)+\alpha\delta\Big)^2}\Bigg]^2 - \frac{4\alpha\beta(1-\kappa-\delta)^2}{\Big(\kappa+\alpha(1-\delta)\Big)^2\Big((1-\kappa)+\alpha\delta\Big)^2}}}{2}
\end{align}
Thus,
\begin{align}
    \lambda_{VBD}^{max} = \frac{\frac{\kappa^2+\alpha(1-\delta)^2}{\Big(\kappa+\alpha(1-\delta)\Big)^2} + \frac{\beta\Big((1-\kappa)^2+\alpha\delta^2\Big)}{\Big((1-\kappa)+\alpha\delta\Big)^2} + \sqrt{\Bigg[\frac{\kappa^2+\alpha(1-\delta)^2}{\Big(\kappa+\alpha(1-\delta)\Big)^2} + \frac{\beta\Big((1-\kappa)^2+\alpha\delta^2\Big)}{\Big((1-\kappa)+\alpha\delta\Big)^2}\Bigg]^2 - \frac{4\alpha\beta(1-\kappa-\delta)^2}{\Big(\kappa+\alpha(1-\delta)\Big)^2\Big((1-\kappa)+\alpha\delta\Big)^2}}}{2}.
\end{align}
We define Vulnerability in VBD ($\nu^{VBD}$) as $\sqrt{\lambda_{VBD}^{max}}$. Thus, the expression for overall vulnerability of the system is:
\begin{align}\label{VBD}
    \nu^{VBD} = \sqrt{\frac{\Tilde{\Theta}_h(\kappa,\delta) + \beta\Tilde{\Theta}_l(\kappa,\delta) + \sqrt{\Big(\Tilde{\Theta}_h(\kappa,\delta) + \beta\Tilde{\Theta}_l(\kappa,\delta) \Big)^2 - \alpha\beta\Big(\Tilde{\Theta}_{hl}(\kappa,\delta)\Big)^2}}{2},}
\end{align}
where,
\begin{itemize}
    \item hub-vulnerability: $\Tilde{\Theta}_h(\kappa,\delta)=\frac{\kappa^2+\alpha(1-\delta)^2}{\Big(\kappa+\alpha(1-\delta)\Big)^2}$,
    \item leaf-vulnerability: $\Tilde{\Theta}_l(\kappa,\delta)=\frac{\Big((1-\kappa)^2+\alpha\delta^2\Big)}{\Big((1-\kappa)+\alpha\delta\Big)^2}$,
    \item cross-vulnerability: $\Tilde{\Theta}_{hl}(\kappa,\delta)=\frac{2(1-\kappa-\delta)}{\Big(\kappa+\alpha(1-\delta)\Big)\Big((1-\kappa)+\alpha\delta\Big)}$.
\end{itemize}

\newpage
\section{Limiting Cases}
\subsection{Contagion dynamics in ABD}
The vulnerability equation in ABD Eq.\ref{ABD} is:
\begin{align}
    \nu^{ABD} = \frac{\Theta_h(\kappa,\delta) + \frac{\Theta_l(\kappa,\delta)}{\gamma} + \sqrt{\Big(\Theta_h(\kappa,\delta) + \frac{\Theta_l(\kappa,\delta)}{\gamma} \Big)^2 - \frac{\alpha}{\gamma}\Big(\Theta_{hl}(\kappa,\delta)\Big)^2}}{2},
\end{align}
where,
\begin{itemize}
    \item hub-vulnerability: $\Theta_h(\kappa,\delta)=\kappa^2+\alpha(1-\delta)^2$,
    \item leaf-vulnerability: $\Theta_l(\kappa,\delta)=(1-\kappa)^2+\alpha\delta^2$,
    \item cross-vulnerability: $\Theta_{hl}(\kappa,\delta)=2(1-\kappa-\delta)$.
\end{itemize}
\subsubsection{Hub-vulnerability} 
\begin{itemize}
    \item \textbf{Minimum:} When $\kappa=0$ and $\delta=1$, all hub individuals migrate to the leaf, and all leaf individuals stay at the leaf, hub's vulnerability is 0.
    \item \textbf{Maximum:} When $\kappa=1$ and $\delta=0$, all hub individuals stay at the hub, and all leaf individuals migrates to the hub, hub's vulnerability reaches a maximum value of $1+\alpha$.
\end{itemize}
\subsubsection{Leaf-vulnerability} 
\begin{itemize}
    \item \textbf{Minimum:} When $\kappa=1$ and $\delta=0$, all hub individuals stay at the hub, and all leaf individuals migrates to the hub, leaf's vulnerability is 0.
    \item \textbf{Maximum:} When $\kappa=0$ and $\delta=1$, all hub individuals migrate to the leaf, and all leaf individuals stay at the leaf, leaf's vulnerability reaches a maximum value of $1+\alpha$.
\end{itemize}

\subsubsection{Cross-vulnerability}
\begin{itemize}
    \item \textbf{Minimum:} When $\delta=1-\kappa$, the cross vulnerability is 0. This implies that a trade-off is required to nullify cross-vulnerability. Either decreasing $\kappa$ to reduce outflow from the hub and increasing $\delta$ to restrict inflow from the leaf, or increasing $\kappa$ to restrict the hub population and decreasing $\delta$ to encourage migration from the leaf to the hub, can achieve this.
    \item \textbf{Maximum:} The cross vulnerability term has a maximum value of 2 when both $\kappa$ and $\delta$ are 0 i.e. when all hub population migrates to leaf and all leaf population migrates to hub. However, the cross-vulnerability is scaled by the factor $\frac{\alpha}{\gamma}$ in the overall expression of vulnerability. Since we define hotspots based on population density, which results in $\frac{\alpha}{\gamma}<1$ (i.e. $\frac{n_{leaf}}{a_{leaf}}<\frac{n_{hub}}{a_{hub}}$). Thus, the cross-vulnerability term is always small.
\end{itemize}
\textbf{Hub as the hotspot}:
\\
As observed earlier, depending on the values of $\kappa$ and $\delta$, the vulnerability of the hub and leaf can reach a maximum value of $1+\alpha$ at different points. This implies that mobility dynamics can alter the relative vulnerability of the hub and leaf, potentially reversing their roles as hotspots. In other words, each patch can act as a hotspot under different mobility conditions. However, assuming a significantly smaller $\frac{\alpha}{\gamma}$ ratio (i.e., $\frac{\alpha}{\gamma}<<1$), indicating a significantly larger leaf area, the contact density in the leaf is very low. This suggests that most contagion events occur in the hub, increasing the likelihood of persistent contagion and reinforcing its role as a hotspot.


\newpage
\subsection{Contagion dynamics in VBD}
The vulnerability equation in ABD Eq.\ref{VBD} is:
\begin{align}
    \nu^{VBD} = \sqrt{\frac{\Tilde{\Theta}_h(\kappa,\delta) + \beta\Tilde{\Theta}_l(\kappa,\delta) + \sqrt{\Big(\Tilde{\Theta}_h(\kappa,\delta) + \beta\Tilde{\Theta}_l(\kappa,\delta) \Big)^2 - \alpha\beta\Big(\Tilde{\Theta}_{hl}(\kappa,\delta)\Big)^2}}{2},}
\end{align}
where,
\begin{itemize}
    \item hub-vulnerability: $\Tilde{\Theta}_h(\kappa,\delta)=\frac{\kappa^2+\alpha(1-\delta)^2}{\Big(\kappa+\alpha(1-\delta)\Big)^2}$,
    \item leaf-vulnerability: $\Tilde{\Theta}_l(\kappa,\delta)=\frac{\Big((1-\kappa)^2+\alpha\delta^2\Big)}{\Big((1-\kappa)+\alpha\delta\Big)^2}$,
    \item cross-vulnerability: $\Tilde{\Theta}_{hl}(\kappa,\delta)=\frac{2(1-\kappa-\delta)}{\Big(\kappa+\alpha(1-\delta)\Big)\Big((1-\kappa)+\alpha\delta\Big)}$.
\end{itemize}

Unlike ABDs, VBDs are confined within patches and follow a mean-field assumption that vectors are restricted to their respective patches and can only interact with individuals present within those patches, regardless of whether they are residents or incoming populations.  Consequently, vulnerability in each patch cannot be completely eliminated if vectors are present, as they cannot migrate to other patches.  Mathematically, this arises because contact density ($f$) in ABD is directly proportional to effective population ($\frac{n^{eff}}{area}$), while in VBD, it is inversely proportional to effective population ($\frac{vector-population}{n^{eff}}$). Consequently, the contact density/effective population term in the vulnerability equation cancels out the effective population term in ABD, but results in $\frac{1}{(n^{eff})^2}$ in VBD.

\subsubsection{Hub-vulnerability} 
\begin{itemize}
    \item \textbf{Minimum:} When $\kappa = 1-\delta$ i.e. when either decreasing $\kappa$ to reduce outflow from the hub and increasing $\delta$ to restrict inflow from the leaf, or increasing $\kappa$ to restrict the hub population and decreasing $\delta$ to encourage migration from the leaf to the hub, hub's vulnerability is minimized to a value of $\frac{1}{1+\alpha}$. 
    \item \textbf{Maximum:}  In contrast, When $\kappa\approx 0$ and $\delta<1$, most hub individuals migrates to the leaf, and some leaf individuals migrates to the hub, hub's vulnerability reaches a maximum value of $\frac{1}{\alpha}$ (provided $\alpha\leq 1$).
\end{itemize}

\subsubsection{Leaf-vulnerability} 
\begin{itemize}
    \item \textbf{Minimum:} When $\kappa = 1-\delta$ i.e. when either decreasing $\kappa$ to reduce outflow from the hub and increasing $\delta$ to restrict inflow from the leaf, or increasing $\kappa$ to restrict the hub population and decreasing $\delta$ to encourage migration from the leaf to the hub, leaf's vulnerability is minimized to a value of $\frac{1}{1+\alpha}$. 
    \item \textbf{Maximum:}  In contrast, When $\kappa\approx 1$ and $\delta>0$, most hub individuals stay the hub, and some leaf individuals stay at the leaf, leaf's vulnerability reaches a maximum value of $\frac{1}{\alpha}$ (provided $\alpha\leq 1$).
\end{itemize}
Similar to ABDs, the contribution of the leaf to the vulnerability of the hub for VBD) is scaled by a spatial factor. However, for VBDs, this spatial factor is $\beta$, the proportion of the leaf's vector population relative to the hub.

\subsubsection{Cross-vulnerability}
\begin{itemize}
    \item \textbf{Minimum:} When $\delta=1-\kappa$, the cross vulnerability is 0. This implies that a trade-off is required to nullify cross-vulnerability. Either decreasing $\kappa$ to reduce outflow from the hub and increasing $\delta$ to restrict inflow from the leaf, or increasing $\kappa$ to restrict the hub population and decreasing $\delta$ to encourage migration from the leaf to the hub, can achieve this.
    \item \textbf{Maximum:} The cross vulnerability term has a maximum value of $\frac{2}{\alpha}$ (provided $\alpha\leq 1$) when both $\kappa$ and $\delta$ are 0 i.e. when all hub population migrates to leaf and all leaf population migrates to hub. However, the cross-vulnerability is scaled by the factor $\alpha\beta$ in the overall expression of vulnerability. Since both $\alpha$ and $\beta$ are typically small, the cross-vulnerability term is generally negligible.
\end{itemize}

\newpage
\section{Epidemic vulnerability in ABD}
\begin{figure}[H]
    \centering
    \includegraphics[width=16cm]{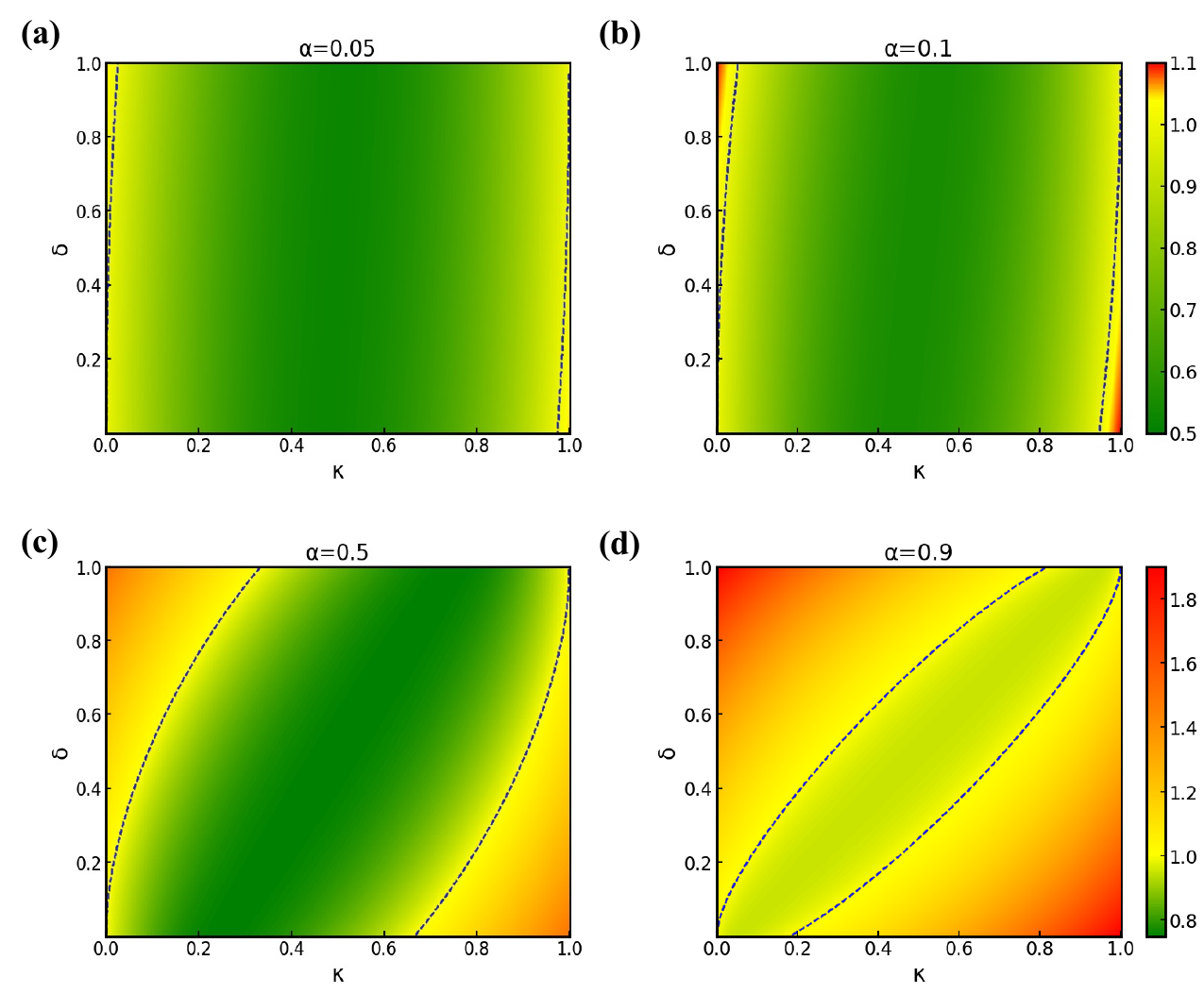}
    \caption{\textbf{Epidemic vulnerability in ABD:} Contour plots illustrating the dynamics of vulnerability in vector-borne disease (ABD) for different values of the parameter $\alpha$. The blue dotted contour line indicates the threshold where vulnerability equals 1. In all plots, $\gamma=1$.}
    \label{fig:S5}
\end{figure}

\newpage
\section{Epidemic vulnerability in VBD}
\begin{figure}[H]
    \centering
    \includegraphics[width=16cm]{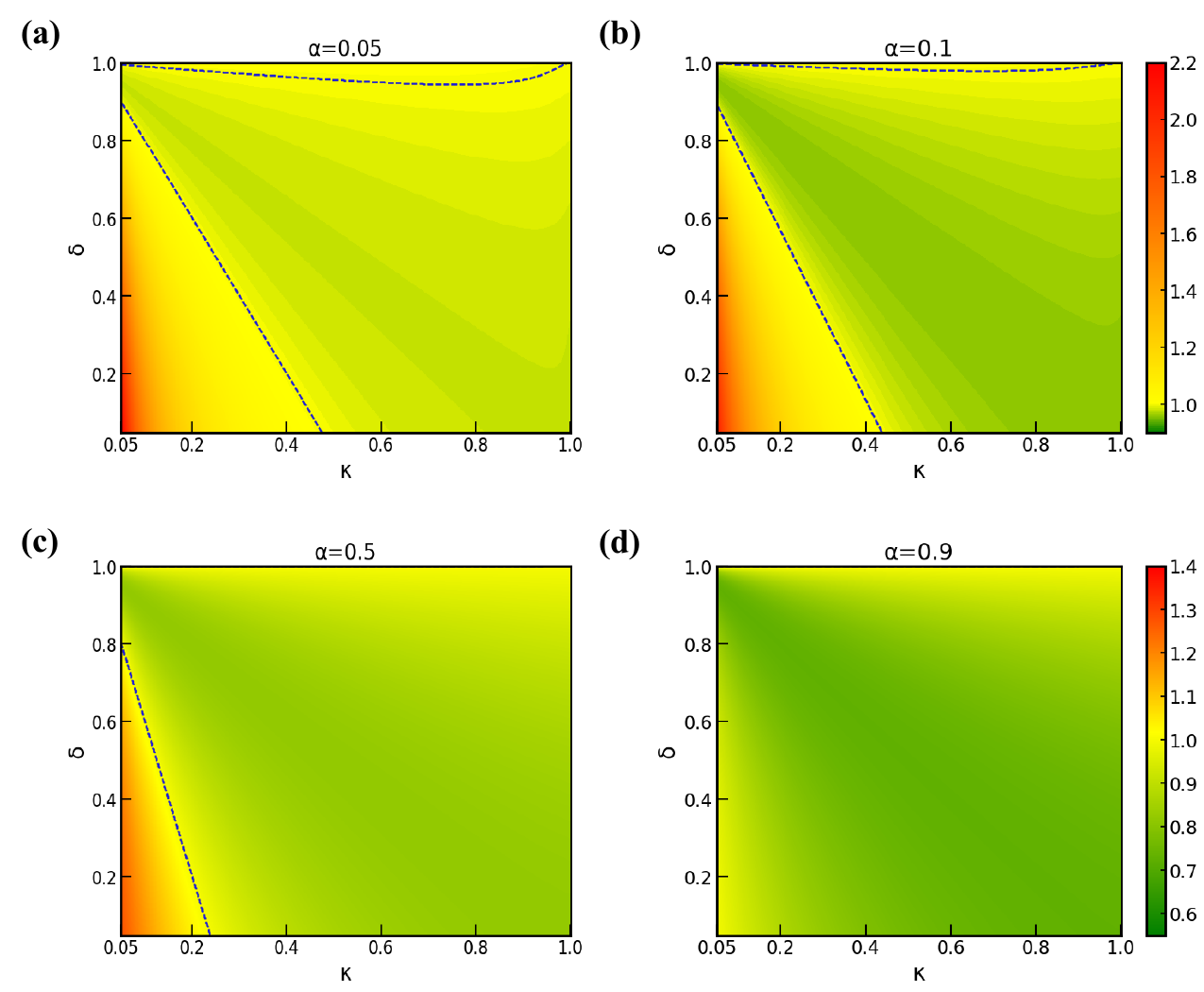}
    \caption{\textbf{Epidemic vulnerability in VBD:} Contour plots illustrating the dynamics of vulnerability in vector-borne disease (VBD) for different values of the parameter $\alpha$. The blue dotted contour line indicates the threshold where vulnerability equals 1. In all plots, $\beta=0.01$.}
    \label{fig:S5}
\end{figure}
\newpage
\section{Application of NPI Strategies to Cali}

For Strategy I, values for $\delta$ were randomly sampled from a uniform distribution between 0 and 1, while $\kappa$ was determined using the constraint $\kappa = 1 - \delta$. These values of $\kappa$ were redistributed equally among hotspot patches, with $1 - \kappa$ allocated as the total outward flow from hotspots to suburban patches. Similarly, $\delta$ was distributed equally among suburban patches and within suburban patches, with $1 - \delta$ representing the total outward flow from suburban patches to hotspot patches. This redistribution ensured that the modified mobility matrix remained normalized. The ratio of the modified vulnerability to the baseline vulnerability was then calculated for evaluation.

For Strategy II, the desired values of $\delta$ ($\frac{\gamma}{\gamma+1}$) and $\kappa$ ($\frac{1}{\gamma+1}$) were first calculated. To account for patch-level variability, values were randomly sampled from Gaussian distributions centered on these desired values, with a standard deviation of 0.2. This allowed for realistic variability while adhering to the general parameter guidelines.

The results indicated that Strategy I was beneficial for both ABDs and VBDs, effectively reducing vulnerability. However, it was slightly less effective than Strategy II for ABDs, highlighting the latter's ability to optimize mobility parameters more precisely and adapt to the distinct transmission dynamics of airborne diseases.